\documentclass[11pt]{article}

\usepackage{amssymb,latexsym,amssymb,amsmath}
\usepackage{slashed}
\usepackage[latin1]{inputenc}
\usepackage{subfigure}
\usepackage{nicefrac}
\usepackage{bbm}
\usepackage[stable]{footmisc}
\usepackage{fancyheadings}
\usepackage{fancybox}
\usepackage{ifthen}
\usepackage[errorshow]{tracefnt}
\usepackage{multirow}
\usepackage{rangecite}

\newcommand{\nn}{\nonumber}

\makeatletter
\@addtoreset{equation}{section}
\makeatother

\def\a {a}
\def\b {b}
\def\g {g}

\newcommand{\be}{\begin{equation}}
\newcommand{\ee}{\end{equation}}

\begin{document}

\begin{titlepage}

\rightline{\small LMU-ASC 35/12}
\rightline{\small MPP-2012-93}

\vskip 2.8cm   
  \vspace*{\stretch{1}}
  \begin{center}
    \Large Heterotic domain wall solutions and $SU(3)$ structure manifolds
    
  \end{center}
  \vspace*{\stretch{2}}
  \begin{center}
    \begin{minipage}{\textwidth}
      \begin{center}
        James Gray${}^{1}$,
        Magdalena Larfors${}^{1}$ and
        Dieter L\"ust${}^{1,2}$
      \end{center}
    \end{minipage}
  \end{center}
  \vspace*{1mm}
  \begin{center}
    \begin{minipage}{\textwidth}
      \begin{center}
        ${}^1$Arnold-Sommerfeld-Center for Theoretical Physics,\\
        Department f\"ur Physik, Ludwig-Maximilians-Universit\"at, M\"unchen,\\
        Theresienstrasse 37, 80533 M\"unchen, Germany\\[0.2cm]
        ${}^2$Max-Planck-Institut f\"ur Physik - Theorie,\\
        F\"ohringer Ring 6, 80805 M\"unchen, Germany
      \end{center}
    \end{minipage}
  \end{center}
  \vspace*{\stretch{1}}
  \begin{abstract}
    \normalsize 

We examine compactifications of heterotic string theory on manifolds with $SU(3)$ structure. In particular, we study ${\cal N} = 1/2$ domain wall solutions which correspond to the perturbative vacua of the $4D$, ${\cal N} =1$ supersymmetric theories associated to these compactifications. We extend work which has appeared previously in the literature in two important regards. Firstly, we include two additional fluxes which have been, heretofore, omitted in the general analysis of this situation. This allows for solutions with more general torsion classes than have previously been found. Secondly, we provide explicit solutions for the fluxes as a function of the torsion classes. These solutions are particularly useful in deciding whether equations such as the Bianchi identities can be solved, in addition to the Killing spinor equations themselves. Our work can be used to straightforwardly decide whether any given $SU(3)$ structure on a six-dimensional manifold is associated with a solution to heterotic string theory. To illustrate how to use these results, we discuss a number of examples taken from the literature.

  \end{abstract}
  \vspace*{\stretch{5}}
  \begin{minipage}{\textwidth}
    \underline{\hspace{5cm}}
    \\
    \noindent {\small{\texttt{james.gray@physik.uni-muenchen.de, magdalena.larfors@physik.uni-muenchen.de, dieter.luest@lmu.de}}}
      \end{minipage}
\end{titlepage}

\tableofcontents 

\section{Introduction}

G-structures on Riemannian manifolds play an important role in all of the standard approaches to string phenomenology, see \cite{Grana:2005jc} for reviews. In heterotic string theory, for example, compactification on a six-dimensional manifold admitting an $SU(3)$ structure naturally leads to an ${\cal N}=1$ theory in four dimensions. In the presence of such a structure there is a single no-where vanishing six-dimensional spinor, invariant under the $SU(3)$ structure group, in terms of which the ten-dimensional fermions and supersymmetry parameters can be decomposed. Frequently, in dealing with heterotic compactifications, one specializes further and considers a manifold admitting not only an $SU(3)$ structure but one which is free of intrinsic torsion. There are several advantages to dealing with such Calabi-Yau threefolds. Firstly, Calabi-Yau threefolds are the only manifolds of $SU(3)$ structure which are also algebraic varieties. The use of algebraic geometry allows these manifolds to be studied in their hundreds of thousands, and gauge field configurations over them give rise to even greater numbers of solutions of string theory (see \cite{Bouchard:2005ag} for some recent work). A second advantage to working with Calabi-Yau threefolds is that they give rise to a perturbative ${\cal N}=1$ Minkowski {\it vacuum} to the associated ${\cal N}=1$ theory \cite{Candelas:1985en}. This leads to many practical advantages in discussing string cosmology and phenomenology.

There are, however, disadvantages to compactifying heterotic string theory on Calabi-Yau threefolds. Traditionally moduli stabilization has proven to be difficult in such backgrounds, although progress has been made in this direction recently (see \cite{hetCYstable} for example). Generalizing to $SU(3)$ structures with intrinsic torsion gives rise to additional contributions to the ${\cal N}=1$ superpotential of the four-dimensional theory which can help greatly in this regard. This, along with the fact that these compactifications are, in some sense, a more general case, makes the study of $SU(3)$ manifolds of great interest. There are, however, disadvantages to moving away from Calabi-Yau threefolds. In addition to the loss of direct applications of the tool of algebraic geometry, in general such compactifications do not have a perturbative ${\cal N}=1$ Minkowski vacuum. 

A subset of the possible intrinsic torsions which can be associated to an $SU(3)$ structure do lead to Minkowski vacua. The Strominger system \cite{Strominger:1986uh}, which in modern language corresponds to a particular restriction on the torsion classes of the $SU(3)$ structure \cite{Cardoso:2002hd}, describes the general case in which the ${\cal N}=1$ theory obtained by dimensional reduction has a ${\cal N}=1$ perturbative Minkowski vacuum. This may seem, therefore to be the most general system of interest. However, as has been discussed a number of times, more general situations can be of equal import (for related work see \cite{4Dhalfflat,Louis:2009dq,Lust:2004ig}).

The crucial observation is that, phenomenologically, one wishes to have a vacuum which is (close to) Minkowski after {\it all} effects, both non-perturbative and perturbative, have been taken into account. Non-perturbative corrections to the superpotential are required in many moduli stabilization scenarios, and so to demand the existence of  a Minkowski vacuum in their absence is clearly too restrictive. One could well have a situation where the perturbative vacuum is, for example, an ${\cal N}=1/2$ domain wall solution in four dimensions,\footnote{By ${\cal N}=1/2$ we mean that the vacuum preserves two supercharges rather than four. See the discussion around equation \eqref{spinoransatz} for more detail.} but where the true vacuum, once effects such as gaugino condensation and membrane instantons are included, is (close to) Minkowski and maximally supersymmetric. It is clear that such a situation requires fine tuning. The balancing of perturbative against non-perturbative effects in this fashion requires either the perturbative effects to be anomalously small, in a fashion which is morally equivalent to what occurs in the KKLT scenario \cite{Kachru:2003aw}, or the non-perturbative effects to be anomalously large, similarly to what occurs in LARGE volume compactifications \cite{Balasubramanian:2005zx}. Whether one has enough freedom to, say, tune the perturbative running in the domain wall to be sufficiently small to be balanced by non-perturbative effects is a case dependent question. Indeed, if the running can not be tuned to be small in this manner, there is no reason to expect a separation in scales between the curvatures seen in four dimensions and the size of the compact manifold - rendering a discussion in lower dimensions meaningless. Nevertheless, the above two examples from type IIB string compactifications do show the importance of such possibilities.\footnote{Note that due to the no-scale structure of type IIB compactifications one does not have a perturbative domain wall solution in the absence of non-perturbative effects in the KKLT scenario (one does have domain wall solutions in type IIA \cite{Haack:2009jg,Mayer:2004sd}, as well as domain walls mediating between perturbative type IIB vacua, see e.g. \cite{Johnson:2008kc}). Nevertheless, to obtain the final vacuum one does have to balance perturbative against non-perturbative effects in the superpotential leading to the fine tuning of the perturbative superpotential.}

In this paper, we study  solutions of heterotic string theory which are compactifications on manifolds of $SU(3)$ structure and which exhibit a domain wall in four dimensions. These solutions, which preserve ${\cal N}=1/2$ of the four-dimensional supersymmetries should be thought of as perturbative vacuum solutions of the associated ${\cal N}=1$ compactified theories.\footnote{One could also consider ${\cal N}=0$ solutions as in \cite{Held:2010az}.} In fact, knowledge of these solutions is a necessary prerequisite to computing the four-dimensional theory itself, which describes massless and low energy fluctuations about such configurations. We begin by providing a general formalism for analyzing such solutions given the torsion classes of the six-dimensional $SU(3)$ structure in question. In particular we solve the relevant Killing spinor equations to give the supergravity fields as a function of these torsion classes, in addition to stating the constraints on the torsion classes that are necessary for such a solution. This work generalizes that of \cite{Lukas:2010mf} by adding in all fluxes allowed by the symmetry requirements, some of which were omitted in that work. In addition, we provide solutions for the fluxes and forms characterizing the domain wall dependence of the system as a function of the torsion classes involved. The nature of these general solutions will be described in detail in section 4. To give an example of the type of result obtained, and to stress how explicit a form it takes, the following is the solution for the three types of Neveu--Schwarz flux present in the system, in terms of the forms characterizing the manifold of $SU(3)$ structure

\vspace{0.1cm}

\begin{equation*}
\begin{aligned}
f &= \alpha_{1+} + 3 e^{\Delta} W_{1-} \\ \nn
H_y &= e^{\Delta} (-f -2 W_{1-}) J - e^{\Delta} W_{2-} + \frac{1}{2} e^{\Delta} ( (2 W_4 - W_5)\llcorner \overline{\Omega} + \textnormal{c.c.}) \\ \nn
H &= -\frac{1}{2} e^{-\Delta} \phi' \Omega_{+} + (\frac{7}{8} f + \frac{3}{2} W_{1-}) \Omega_{-} + *((3W_4-2 W_{5+}) \wedge J - W_3 + e^{-\Delta} \alpha_3) \; . \\  \;
\end{aligned}
\end{equation*}
From these expressions, it is clear that almost all torsion classes can be balanced by appropriate choices of fluxes. Indeed, we only find one direct necessary condition on the torsion classes, namely that $W_4$ must be exact.

Explicit solutions for the supergravity fields, such as these, are important as they allow us to check the Bianchi identities and form field equations of motion - which must be checked independently of the Killing spinor equations. The formalism which we present provides a general set of rules which can be applied to any given manifold with $SU(3)$ structure to see if it allows a heterotic ${\cal N}=1/2$ domain wall solution. Once this general formalism has been explained we demonstrate how to apply it to a number of cases. These include the $SU(3)$ structures explicitly constructed on compact toric threefolds in \cite{Larfors:2010wb}, as well as a number of other cases taken from the literature.

Mathematically, the combination of the domain wall direction and the compact six-dimensional manifolds in these solutions provide a generalization of Hitchin flow \cite{Hitchin:2001rw}. Instead of a six-dimensional $SU(3)$ structure manifold varying over an extra direction to form a manifold with $G_2$ holonomy, a more general six-dimensional structure is flowing to form a manifold with $G_2$ structure. Hitchin flow itself can, of course, be recovered as a special case of this analysis, when all flux is set to zero. For related discussions see \cite{Lukas:2010mf,Mayer:2004sd,Jeschek:2004wy}.

The rest of this paper is structured as follows. In section 2 we present the ansatz we shall work with for the supergravity fields and derive the equations which our domain wall configurations must satisfy. In section 3 we present a characterization of both the possible supergravity fluxes, and the domain wall dependence of the various forms in the theory. In section 4 we analyse this system and solve for the supergravity fields and domain wall variation. Sections 5, 6 and 7 contain several examples of the application of this analysis to geometries found in the literature. The first example is that of Calabi-Yau compactifications where the back reaction of supergravity flux is purely in the domain wall direction and does not deform the compact space. The second example is an analysis of coset manifolds admitting an $SU(3)$ structure. The final example is based upon recent work explicitly defining $SU(3)$ structures on smooth compact toric varieties \cite{Larfors:2010wb}. We make a few concluding remarks in section 8, and two technical appendices are provided in order to make this paper self contained.

\section{Ansatz and equation system} \label{sec:eqnsys}

We wish to study solutions of heterotic string theory which take the form of a warped product of a manifold admitting an $SU(3)$ structure and a four-dimensional domain wall. In addition we will require these solutions to preserve a number of supercharges corresponding to ${\cal N}=1/2$ supersymmetry in four dimensions. Given this, the first equations for us to study are the supersymmetry variations of the gravitino and dilatino in ten dimensions (solutions in this paper will be to lowest order in $\alpha'$ and thus will not involve gauge fields)
\begin{eqnarray}\label{ks}
  \delta \psi_M &=& \left(\triangledown_M + \frac{1}{8} \hat{{\cal H}}_M \right) \epsilon \\ \nonumber
  \delta \lambda &=& \left( \slashed{\triangledown} \hat{\phi} + \frac{1}{12} \hat{{\cal H}} \right) \epsilon \; .
\end{eqnarray}
Here, hatted quantities indicate ten-dimensional supergravity fields, $\hat{{\cal H}}_M = \hat{H}_{MNP} \Gamma^{NP}$ and $\hat{{\cal H}} = \hat{H}_{MNP} \Gamma^{MNP}$ and we are searching for backgrounds such that these variations are zero for two supercharges $\epsilon$. Given our interest in domain wall solutions, we make the following ansatz for the metric
\begin{eqnarray} \label{metric}
ds_{10}^2 = e^{2 A (x^m)} \left( ds_3^2 + e^{2 \Delta(x^u)} dy dy + g_{uv}(x^m) dx^u dx^v\right) \; .
\end{eqnarray}
Here $ds_3^2$ corresponds to any maximally symmetric space for the world volume of the domain wall, the coordinates $x^m$ include the six-dimensional coordinates and the domain wall direction and the coordinates $x^u$ are those on the manifold of $SU(3)$ structure. The coordinate $y$ is normal to the domain wall and we denote coordinates on the domain wall world volume by $x^{\alpha}$. In order to preserve the maximal symmetry of the world volume we make the ansatzes  $\partial_{\alpha} \hat{\phi} =0$ and $\hat{H}_{\alpha \beta \gamma} = f \epsilon_{\alpha \beta \gamma}$ where $f$ is some function and $\epsilon_{\alpha \beta \gamma}$ is the volume form on the world volume. We also allow $\hat{H}_{ymn}$ to be arbitrary and insist that $H_{\alpha m n}=H_{\alpha \beta n}=0$ , to be consistent with our symmetry requirements. The system considered by Lukas and Matti \cite{Lukas:2010mf} can be recovered by taking the special case where the domain wall is Minkowski space, and by setting $H_{y m n}$ and $f$ to zero.

To analyse the Killing spinor equation \eqref{ks}, we must make an ansatz for the spinor $\epsilon$ which is compatible with the form of our metric \eqref{metric}. Following the notation of \cite{Lukas:2010mf}, we write
\begin{eqnarray}\label{spinoransatz}
\epsilon(x^{\alpha}, x^m) = \rho(x^{\alpha}) \otimes \eta(x^m) \otimes \theta  \; .
\end{eqnarray}
Here $\rho$ is the standard covariantly constant spinor on the Minkowski or anti de Sitter (AdS) world volume of the domain wall, $\eta$ is a seven-dimensional Majorana spinor on the seven-dimensional space composed of the normal direction to the domain wall and the compact six-dimensional manifold and $\theta$ is an eigenvector of the third Pauli matrix. The spinor $\rho$ has two components and corresponds to the two conserved supercharges which we wish to obtain in four or three dimensions. For later use we also note that we will sometimes split $\eta$ up in terms of two six-dimensional spinors of definite chirality 
\be
\eta=\frac{1}{\sqrt{2}} (\eta_+ + \eta_-) \; .
\ee

Given the above ansatzes, \eqref{metric} and \eqref{spinoransatz}, one may compute the components of the gravitino variation \eqref{ks} in the domain wall direction. One then finds that $\partial_m A=0$. Since this warp factor is therefore a constant, its value can be absorbed into a coordinate redefinition, and so we set $A=0$ for the remainder of our analysis. With these definitions and initial simple steps dealt with, we now proceed onto a more detailed analysis of the remaining conditions in equations \eqref{ks}.

\vspace{0.1cm}
 
As mentioned in the introduction, we can combine the normal direction to the domain wall and the six manifold with $SU(3)$ structure into a non-compact seven manifold with $G_2$ structure. Let us rephrase the Killing spinor equations \eqref{ks} in terms of the forms of this $G_2$ structure. We have the following definitions
\begin{eqnarray}
  \varphi_{mnp} = - i \eta^{\dagger} \gamma_{mnp} \eta \;,\;\; \Phi_{mnpq} = \eta^{\dagger} \gamma_{mnpq} \eta \; .
\end{eqnarray} 
The Killing spinor equations \eqref{ks} then imply the following
relations
\begin{eqnarray} \label{first} \triangledown_m \varphi_{npq} &=&
  \frac{3}{2} \hat{H}_{ms [n} \varphi_{pq]}^s \\ \label{second}
  \triangledown_m \Phi_{npqr} &=& -2 \hat{H}_{ms[n} \Phi_{pqr]}^s
  \\ \label{third} \triangledown_{[m} \hat{\phi} \Phi_{npqr]} &=&
  \hat{H}_{s[mn} \Phi_{pqr]}^s \\ \label{fourth} 4 \triangledown_{[m}
  \hat{\phi} \varphi_{npq]} &=& -3 \hat{H}_{v[nm} \varphi_{pq]}^v +
  \frac{1}{12} \epsilon_{nmpqrst} \hat{H}^{rst} + \frac{1}{2}
  \Phi_{mnpq} f \\ \label{fifth}
  \epsilon_{mnpqrst} \triangledown^t \hat{\phi} &=&-10 \hat{H}_{[mnp} \varphi_{qrs]} \\ \label{sixth}
  0 &=& \frac{1}{2} \epsilon_{mnpqrst} f - 35 \hat{H}_{[mnp}
  \Phi_{qrst}] \; .
\end{eqnarray} 
Combining equations \eqref{first} and \eqref{fourth}, and equations \eqref{second} and \eqref{third}, and taking equations \eqref{fifth} and \eqref{sixth} without change, we may write the following
\begin{eqnarray} \label{7dformeqns} d_7 \varphi &=& 2 d_7 \hat{\phi}
  \wedge \varphi - *_7 \hat{H} - \Phi f \;, \\ \nonumber
  d_7 \Phi &=& 2 d_7 \hat{\phi} \wedge \Phi \;, \\ \nonumber
  *_7 d_7 \hat{\phi} &=& - \frac{1}{2} \hat{H} \wedge \varphi \;, \\ \nonumber
  0 &=& \frac{1}{2} *_7 f -\hat{H} \wedge \Phi \; .
\end{eqnarray}
Note that these expressions reduce to those found in \cite{Lukas:2010mf} when you set the extra fluxes we have included to zero.

In rewriting the system in this manner the seven dimensional dependence of the killing spinor and metric has been encoded in the $x^m$ dependence of the forms of the $G_2$ structure. We can now rewrite these equations again, in a language that separates the six-dimensional $SU(3)$ structure from the domain wall direction.\footnote{The following
  relationships can be useful in passing between the six and seven-dimensional versions of the equations we are studying
\begin{eqnarray}
  *_7(\alpha_p^{(6)}) &=& e^{\Delta} dy \wedge *(\alpha_p^{(6)}) \\
  *_7(dy \wedge \alpha_{p-1}^{(6)}) &=& e^{-\Delta} (-1)^{p-1}  * (\alpha_{p-1}^{(6)}) \ .
\end{eqnarray}} The three form and two form of the $SU(3)$ structure can be written
\begin{eqnarray}
\Omega_{uvw} = \eta_+^{\dagger} \gamma_{uvw} \eta_- \;\;\;,\;\;\; J_{uv} = \mp \eta_{\pm}^{\dagger} \gamma_{uv} \eta_{\pm} \; .
\end{eqnarray} 
where $\eta_{\pm}$ are the components of the seven-dimensional spinor $\eta$ of definite six-dimensional chirality. It is straightforward to show that $J$ and $\Omega$ thus defined obey the orthogonality relation
\be \label{eq:orth}
J \wedge \Omega = 0 \; .
\ee
Given these definitions, and our metric ansatz \eqref{metric}, we may relate the forms of the $G_2$ and $SU(3)$ structures \cite{Chiossi:2002tw}
\begin{eqnarray} \label{def1}
\varphi &=& e^{\Delta} dy \wedge J + \Omega_- \\ \label{def2}
\Phi &=& e^{\Delta} dy \wedge \Omega_+ + \frac{1}{2} J \wedge J \; .
\end{eqnarray}
The equations \eqref{7dformeqns} determining the seven-dimensional $G_2$ structure can then be written 
\begin{eqnarray} \label{1}
J \wedge dJ &=& J \wedge J \wedge d \hat{\phi} \\ \label{2}
dJ &=& e^{-\Delta} \Omega'_- - 2 e^{-\Delta} \hat{\phi}' \Omega_- + 2 d \hat{\phi} \wedge J -J \wedge \Theta + * H + f \,\Omega_+ \\ \label{3}
d \Omega_+ &=& e^{-\Delta} J \wedge J' - e^{-\Delta} \hat{\phi}' J \wedge J +  2 d \hat{\phi} \wedge \Omega_+ + \Omega_+ \wedge \Theta \\ \label{4}
d \Omega_- &=& 2 d \hat{\phi} \wedge \Omega_- - e^{-\Delta} * H_y - \frac{1}{2} f J \wedge J \\
\label{5}
0 &=& \frac{1}{2} * f - \Omega_+ \wedge H -e^{-\Delta} \frac{1}{2} H_y \wedge J \wedge J   \\
e^{-\Delta} * \hat{\phi}' &=& -\frac{1}{2} H \wedge \Omega_- \label{6}\\
e^{\Delta} * d \hat{\phi} &=& \frac{1}{2} H_y \wedge \Omega_- -\frac{1}{2} e^{\Delta} H \wedge J \label{7} \; .
\end{eqnarray}
In the above $d$ denotes the exterior derivative in six dimensions, and a prime indicates a derivative with respect to $y$, the coordinate normal  to the domain wall. Moreover, $\Theta = d \Delta$, Hodge stars are taken with respect to the six-dimensional metric and $\Omega_+$ and $\Omega_-$ are the real and imaginary parts respectively of the three form $\Omega$. The field $H$ is the three form corresponding to $\hat{H}$ with all of its indices lying on the six-dimensional compact manifold, whereas $H_y$ is a two form corresponding to $\hat{H}$ with its first index pointing in the normal direction to the domain wall.

\vspace{0.1cm}

It is important to note that the forms $J$ and $\Omega$ are not, in general, closed. The degree to which they fail to be so is classified by the torsion classes of the $SU(3)$ structure \cite{Cardoso:2002hd,Chiossi:2002tw}
\begin{eqnarray} \label{dJclasses} dJ &=& -\frac{3}{2}
  \textnormal{Im}(W_1 \overline{\Omega}) + W_4 \wedge J + W_3
  \\ \label{dOclasses} d \Omega &=& W_1 J \wedge J + W_2 \wedge J +
  \overline{W}_5 \wedge \Omega \; .
\end{eqnarray}
The above can be taken as a definition of the torsion classes $W_i$ given that they are also defined to have the primitivity properties
\begin{eqnarray} \label{classesconstraints}
W_3 \wedge J = W_3 \wedge \Omega = W_2 \wedge J \wedge J = 0 \; .
\end{eqnarray}
Moreover, $W_2$ is a complex $(1,1)$ form, $W_3$ is a real $(2,1)+(1,2)$-form, $W_4$ is a real one-form, and $W_5$ is a complex $(1,0)$-form. For a nice discussion of this structure, see \cite{Gurrieri:2002wz}. 

\vspace{0.1cm}

In addition to the Killing spinor equations already discussed, to have a solution to the equations of motion of the system, we must also satisfy the form field equation of motion and Bianchi identity. The components of the ten dimensional Bianchi identity with three indices lying on the domain wall can only be solved by a constant $f$. In terms of the seven-dimensional Bianchi identity, the Bianchi identity reads as $d\hat{H}=0$. In terms of six-dimensional quantities and exterior derivatives, this becomes,
\begin{eqnarray} \label{eq:bi}
d H = 0 \;\;, \;\; d H_y = H' \;.
\end{eqnarray}
The seven-dimensional form field equation of motion, given by $d_7 (*_7 e^{-2\hat{\phi}} \hat{H}) =0 \;$, may be written in terms of six-dimensional quantities as 
\begin{eqnarray} \label{formeom}
d(* e^{-2 \hat{\phi}- \Delta} H_y) =0 \;\;,\;\; (* e^{-2 \hat{\phi} -\Delta} H_y)' + d * (e^{-2 \hat{\phi} + \Delta} H ) =0 \; .
\end{eqnarray}

\vspace{0.1cm}

To conclude this section, we note that, in what follows, we will find it useful to split our equations into those which involve $y$ derivatives and those that do not. In other words, those which explicitly capture how the system evolves in the domain wall direction and those which encapsulate this only implicitly, more directly describing consistency conditions at a single value of $y$. 

\subsubsection*{Consistency at fixed y}

\begin{eqnarray}
  J \wedge d J &=& J \wedge J \wedge d \hat{\phi}  \label{eq1}\\
  d \Omega_- &=& 2 d \hat{\phi} \wedge \Omega_- - e^{-\Delta} * H_y - \frac{1}{2} f J \wedge J  \label{eq2}\\
  0 &=& \frac{1}{2} * f - \Omega_+ \wedge H - \frac{1}{2} e^{-\Delta} H_y \wedge J \wedge J  \label{eq33}\\
  e^{\Delta}* d \hat{\phi} &=& \frac{1}{2} H_y \wedge \Omega_- -\frac{1}{2} e^{\Delta} H \wedge J  \label{eq4}\\
dH &=& 0  \label{eq5} \\ \label{eqn5b}
d(* e^{-2 \hat{\phi}- \Delta} H_y) &=&0  \\ \label{eq6}
df &=& 0
\end{eqnarray}

\subsubsection*{Flow equations}

\begin{eqnarray} \label{floweq1}
  \hat{\phi} ' &=& -\frac{1}{2} e^{\Delta} * (H \wedge \Omega_-) \\ \label{floweq2}
  J \wedge J' &=& e^{\Delta} d \Omega_+  - \frac{1}{2} e^{\Delta} *(H \wedge \Omega_-) J \wedge J - 2 e^{\Delta} d \hat{\phi} \wedge \Omega_+ - e^{\Delta} \Omega_+ \wedge \Theta \\ \label{floweq3}
  \Omega_-' &=& e^{\Delta} dJ  - e^{\Delta}  *(H \wedge \Omega_-) \Omega_- -2 e^{\Delta} d \hat{\phi} \wedge J + e^{\Delta} J \wedge \Theta - * H e^{\Delta} - f e^{\Delta} \Omega_+  \\ \label{floweq4}
H' &=& d H_y \\ \label{floweq5}
 (* e^{-2 \hat{\phi} -\Delta} H_y)' &=&- d * (e^{-2 \hat{\phi} + \Delta} H )  \\ \label{floweq6}
 f' &=& 0 
\end{eqnarray}
Notice these last equations are all of the form ``$y$ derivative $=$ source''.

\section{The most general flux and domain wall dependence}\label{sec:genflux}

In the previous section we have given the equations which must be solved to find ${\cal N}=1/2$ domain wall solutions of heterotic string theory. In this section we detail expansions which can be made, without loss of generality, for the supergravity fields and their derivatives. These expansions are in terms of quantities associated with the $SU(3)$ structure of the six-dimensional compact space and will facilitate the analysis of these equations in the next section.

\subsection{Neveu-Schwarz flux}

We begin by considering the three form field strength $H$ and the two form field strength $H_y$. Manifolds admitting an $SU(3)$ structure are almost complex, and thus any form can be decomposed with respect to its index structure. We can, in complete generality, write $H$ and $H_y$, which are a priori arbitrary three and two forms respectively, in the following manner
\begin{eqnarray} \label{mrH}
H &=& A_{1+} \Omega_+ + A_{1-} \Omega_- + A_{2+} \wedge J + A_{3+} 
\\ \nonumber
H_y &=& B_1 J + B_{2}  + B_{3+}  \; .
\end{eqnarray}
Here $A_{1\pm}, B_1$ are real functions, $A_{2+}$ is the real part of a (1,0)-form, $A_{3+}$ is the real part of a (2,1)-form, $ B_{2}$ is a (1,1)-form and $B_{3+}$ is the real part of a (2,0)-form. These forms can be chosen to obey the primitivity relations
\begin{eqnarray} \label{a31}
A_{3+} \wedge \Omega_{\pm} &=& 0 \\ \label{a32}
A_{3+} \wedge J &=& 0 \\
 B_{2} \wedge J \wedge J &=& 0  \; .
\end{eqnarray} 
In imposing these conditions, we have used the uniqueness of the volume form and the holomorphic top-form, and the freedom to choose $A_{1+} \Omega_+ \wedge \Omega_- = H \wedge \Omega_-$, $A_{1-} \Omega_{-} \wedge \Omega_{+} = H \wedge \Omega_+$, $B_1 J \wedge J \wedge J = H \wedge J \wedge J$ and $A_{2+} =\frac{1}{4} J \llcorner H$ given the initially unspecified nature of $A_{3+}$, $B_2$ and $B_{3+}$.\footnote{Note that $B_{3+} \wedge J \wedge J =0$ is trivially true by index structure arguments.} In addition, this choice of $A_{2+}$ ensures that $J \llcorner A_{3+}=0$.

\vspace{0.1cm}

Given the expansion \eqref{mrH} for $H$ and $H_y$, the six-dimensional Hodge duals of these quantities are readily computed using identities in appendix \ref {app:conv}
\begin{eqnarray}
*H &=& -A_{1+} \Omega_- + A_{1-} \Omega_+ - A_{2-} \wedge J + *A_{3+} 
\\
*H_y &=& \frac{1}{2} B_1 J \wedge J  -  B_{2} \wedge J + *B_{3+} \ ,
\end{eqnarray}
where, as a consequence of $A_{3+} \wedge \Omega_{\pm} = 0$ and $J \llcorner A_{3+}  =0$, 
\begin{eqnarray}
 \Omega_{\pm} \wedge *A_{3+}  = J \wedge *A_{3+} =0 \ .
\end{eqnarray}
 We keep $*B_{3+}$ as it is for future convenience - in the following, we only need to know that $* B_{3+}$ is the real part of a (1,3)-form.

\subsection{Domain wall dependence}

We wish to write down a decomposition for the domain wall dependence ($y$ derivatives) of the forms defining the $SU(3)$ structure, $J$ and $\Omega$, similar to that given for the flux in the proceeding subsection.  In general it may be possible to deform a given $SU(3)$ structure in many ways while preserving the conditions $\Omega \wedge J =0$ and $J \wedge J \wedge J =\frac{3}{4}i \Omega \wedge \overline{\Omega} $. The parameters associated to these deformations, which one might think of as the ``moduli of the $SU(3)$ structure'' in some sense (although of course what constitutes a physical modulus can only be decided once a consistent background solution has been discovered), are what can be allowed to vary with the domain wall direction $y$. This freedom in the $SU(3)$ structure induces a  $y$ dependence of $J$ and $\Omega$. Using arguments similar to those in the previous subsection for the flux, we may, without sacrificing any generality, decompose $J'$ as follows
\begin{eqnarray} \label{jprime}
J' &=& \gamma_1 J + \gamma_{2+} + \gamma_3 \\  
0&=& \gamma_{2+} \wedge J \wedge J = \gamma_3 \wedge J \wedge J \; .
\end{eqnarray}
Here $\gamma_{2+}$ is the real part of a (2,0) form and $\gamma_3$ is
a (1,1) form.

For the three form of the $SU(3)$ structure we may write,
\begin{eqnarray} \label{omegaprime}
  \Omega_-' &=& \alpha_{1+} \Omega_+ + \alpha_{1-} \Omega_- + \alpha_{2+} \wedge J + \alpha_3 \; , \\
  \Omega_+' &=& \beta_{1+} \Omega_+ + \beta_{1-} \Omega_- + \beta_{2+} \wedge J + \beta_3 \; ,\\ 
  0 &=& \Omega_{\pm} \wedge \alpha_3 = J \wedge \alpha_3 \;,  \\ \label{finalomegaprime}
 0 &=& \Omega_{\pm} \wedge \beta_3 = J \wedge \beta_3 \;.
\end{eqnarray}
where we have chosen $\alpha_{2+}=\frac{1}{4} J \llcorner \Omega_-'$, so that $J \llcorner \alpha_3=0$. 

In any given case one can compute the coefficients above, $\alpha, \;\beta$ and $\gamma$, straightforwardly in terms of the $SU(3)$ structure parameters (which one allows to be $y$ dependent). In many cases one may not know all of the possible deformations of the $SU(3)$ structure which is under consideration (there are an infinite number of such deformations). In such a case one can proceed by simply including all deformations which are known and considering a restricted case. In section \ref{case} we consider a set of deformations which are always possible in the case of any known $SU(3)$ structure.

Taking the $y$-derivative of the $SU(3)$ structure conditions we get  consistency conditions, which will automatically be satisfied in any real example. These provide a useful check of calculations and can also be used in performing general analyses without referring to a specific $\Omega$ and $J$:
\begin{eqnarray}
(J \wedge  \Omega) ' = 0 &\implies& \gamma_{2+}  \wedge \Omega = -(i \alpha_{2+} + \beta_{2+}) \wedge J \wedge J\\ \label{consis2}
(J \wedge J \wedge J)' = \frac{3}{4}i (\Omega \wedge \Omega^*)' &\implies& \gamma_1 = \frac{1}{3} (\beta_{1+} + \alpha_{1-}) \; .
\end{eqnarray}

\section{Analysis of the system, consistency relations and flux solution}

In this section we analyse the possible solutions of the equations presented in Section \ref{sec:eqnsys}, using the decompositions presented in Section \ref{sec:genflux}, in complete generality. By substituting our decomposition of the flux and the domain wall dependence of the $SU(3)$ structure forms into the supersymmetry conditions \eqref{1}-\eqref{7}, we derive three different types of conditions. Firstly,  we obtain constraints on the $SU(3)$ structure itself which are required to be satisfied if the system is ever to solve the supersymmetry conditions. Secondly, we obtain solutions for the Neveu-Schwarz flux, the warp factor $\Delta$ and the dilaton in terms of the torsion classes of the $SU(3)$ structure in play. Finally we solve for the domain wall $y$ dependence which is induced in the configuration. The final goal, then, is  to find the supergravity fields, constraints on the $SU(3)$ structure, and domain wall dependence in terms of sums, wedge products and contractions of the torsion classes of a generic $SU(3)$ structure manifold.

\subsection{Supersymmetry conditions: consistency at fixed $y$}

Let us start our analysis with the consistency conditions on the supergravity fields and torsion classes of the $SU(3)$ structure at fixed $y$ - equations \eqref{eq1}-\eqref{eq6}.

\vspace{0.1cm}

The first of these equations, \eqref{eq1}, gives us the following relation
\be \label{sol1}
d \hat{\phi} = W_4 + u_1 \ , \;\;\; \textnormal{where} \;\;\; u_1 \wedge J \wedge J= 0 \; .
\ee
In fact $u_1=0$ as, by dualizing the primitivity condition in \eqref{sol1} one can easily show that any such primitive one form vanishes. Note that, in addition to specifying the exterior derivative of the dilaton, this equation puts a constraint on the torsion class $W_4$ - it must be exact.

\vspace{0.1cm}

Next we consider the wedge product of \eqref{eq2} with $J$. This gives us the relation
\be 
W_{1-} J \wedge J \wedge J = -e^{-\Delta} J \wedge (\frac{1}{2} B_1 J \wedge J) -\frac{1}{2} f J \wedge J \wedge J\;.
\ee
Solving for $B_1$ we then have,
\be \label{sol2}
B_1 =e^{\Delta}(-f -2 W_{1-} )\;.
\ee
We may now go back and analyse the rest of equation \eqref{eq2}. The full equation gives us
\begin{eqnarray} \label{tobreakup}
 W_{2-} \wedge J + W_{5+} \wedge \Omega_- - W_{5-} \wedge \Omega_+ = 2 W_4 \wedge \Omega_- - e^{-\Delta}( - B_2 \wedge J + *B_{3+})  \;,
\end{eqnarray}
where we have used  equations \eqref{sol1} and \eqref{sol2}.

We can analyse equation \eqref{tobreakup} further by breaking it up according to index structure. We have the $(3,1)$ part,
\begin{eqnarray} \label{breakup1}
 W_{5+} \wedge \frac{ \Omega}{2 i} - W_{5-} \wedge \frac{\Omega}{2} = 2 W_4 \wedge \frac{\Omega}{2i} + e^{-\Delta} (* B_{3+})^{(3,1)}  \;\;,
\end{eqnarray}
the $(1,3)$ part,
\begin{eqnarray} \label{breakup2}
 - W_{5+} \wedge \frac{ \overline{\Omega}}{2 i} - W_{5-} \wedge \frac{\overline{\Omega}}{2} = - 2 W_4 \wedge \frac{\overline{\Omega}}{2i} + e^{-\Delta} (* B_{3+})^{(1,3)}  \;\;,
\end{eqnarray}
and finally the $(2,2)$ part (recall that $B_2$ is a (1,1)-form),
\begin{eqnarray} \label{breakup3}
W_{2-} \wedge J = - e^{-\Delta} B_2 \wedge J \;.
\end{eqnarray}

Let us analyse each of these pieces separately and solve for the flux ansatz components where possible. Equations \eqref{breakup1} and \eqref{breakup2} may always be trivially solved for $B_{3+}$ (note that the conjugate nature of the equations is compatible with $B_{3+}$ being real)
\begin{eqnarray} \label{b3pluseqn}
B_{3+}^{(0,2)} &=& \frac{1}{2 i} e^{\Delta} * \left( (2 W_4 - W_{5}) \wedge \overline{\Omega} \right) \\
\Rightarrow B_{3+}^{(0,2)} &=& \frac{1}{2 } e^{\Delta} (2 W_4 -W_5) \llcorner \overline{\Omega} \; .
\end{eqnarray}
To analyse \eqref{breakup3} we simply note that, because $W_{2-} \wedge J \wedge J =0$ (and similarly for $B_2$), we have $W_{2-} \llcorner ( J \wedge J) = -2 W_{2-}^{(1,1)}$. Taking the Hodge dual of equation \eqref{breakup3} then simply leads to the conclusion that,
\begin{eqnarray}
B_2 = - e^{\Delta} W_{2-} \;.
\end{eqnarray}

We continue our analysis with equation \eqref{eq33}. This may be solved in general to give us,
\begin{eqnarray}
A_{1-} = \frac{1}{8} f -\frac{3}{4} e^{-\Delta} B_1\;.
\end{eqnarray}
Using \eqref{sol2}, we then find 
\begin{eqnarray}
A_{1-} = \frac{7}{8} f + \frac{3}{2} W_{1-}\; .
\end{eqnarray}

Next we must analyse equation \eqref{eq4}. Dualizing and using equation \eqref{sol1} we find this implies that,
\begin{eqnarray}
A_{2-}&=& -A_{2+} \llcorner J = - \frac{1}{2} e^{-\Delta} * (B_{3+} \wedge \Omega_-) - W_4  \\
&=& 3 W_4 -2 W_{5+} \;.
\end{eqnarray}
where in the last equation we have used \eqref{b3pluseqn} and expressions from Appendix \ref{mrappy}.

At this stage we are just left with equations \eqref{eq5} and \eqref{eqn5b} from the equations which describe consistency at fixed $y$. Unfortunately the Bianchi identities and form field equation of motion are deceptively complicated to deal with and we therefore defer the solution of these equations to specific cases where simplifications can be made.

\vspace{0.1cm}

We finish this subsection by summarizing what we have learnt.

\begin{itemize}
\item So far, the only constraint on the torsion classes coming from the ${\cal N}=1/2$ supersymmetry relations is that $W_4$ be exact. This is required for a solution.
\item Given such a situation the general solution to these equations for the supergravity fields is
\begin{eqnarray}\label{chargesol}
d \hat{\phi} &=& W_4    \\ \label{thisguy1}
B_1 &=& e^{\Delta}(-f -2 W_{1-}) \\
B_{3+}^{(0,2)} &=&  \frac{1}{2} e^{\Delta}(2 W_4- W_{5})\llcorner \overline{\Omega} \\
B_2 &=& -e^{\Delta} W_{2-}\\ \label{thisguy2}
A_{1-} &=& \frac{7}{8} f +\frac{3}{2} W_{1-} \\  \label{A2mineq}
A_{2-}  &=& 3 W_4 -2 W_{5+} \; .
\end{eqnarray}
From these expressions we can clearly recover the results of \cite{Lukas:2010mf}. In that paper the authors took $H_y=f=0$. Taking all of the $B$'s and $f$ to vanish, we recover from the above that $W_5=2 W_4$, and $W_{1-}=W_{2-}=0$ as claimed in that work.
\item Finally, the Bianchi identities $dH=0$, $df=0$ and the form equation of motion $d(* e^{-2 \hat{\phi} - \Delta} H_y)=0$ must also be solved.
\end{itemize}

\subsection{Supersymmetry conditions: $y$ dependence}

We must now address equations (\ref{floweq1}-\ref{floweq6}), which determine the domain wall dependence of the $SU(3)$ structure and supergravity fields. We begin with equation \eqref{floweq1}. Substituting equation \eqref{mrH} into equation \eqref{floweq1} we obtain
\begin{eqnarray}
A_{1+} = - \frac{1}{2} e^{-\Delta} \hat{\phi}' \; .
\end{eqnarray}
Since $A_{1+}$ has yet to be specified, we can always solve this equation by an appropriate choice of this coefficient in \eqref{mrH}.

We move on to equation \eqref{floweq2}. Firstly, we may take the wedge product of this equation with $J$, which leads to
\begin{eqnarray} \label{mrgamma1}
\gamma_1 = e^{\Delta} ( W_{1+} - 2 A_{1+}) \; .
\end{eqnarray}
Using this, together with the definitions of the torsion classes and our expansion of the fluxes and domain wall dependence, in equation \eqref{floweq2}, we find
\begin{eqnarray} \label{fullJJp}
(\gamma_{2+} + \gamma_3 -e^{\Delta} W_{2+}  )\wedge J =  e^{\Delta}( W_{5+}  -2 W_4  + \Theta)\wedge \Omega_+ + e^{\Delta} W_{5-} \wedge \Omega_- \end{eqnarray}
which we shall now decompose according to almost complex index structure.

We first consider the $(2,2)$ component of expression \eqref{fullJJp}
\begin{eqnarray}
(\gamma_3 - e^{\Delta} W_{2+}) \wedge J =0 \; .
\end{eqnarray}
Clearly this expression tells us that,
\begin{eqnarray} \label{gamma3sol}
\gamma_3 = e^{\Delta} W_{2+} + u_2 \;\; \textnormal{where} \;\; J \wedge u_2 =0 \;.
\end{eqnarray}
By dualizing the primitivity condition in \eqref{gamma3sol}, as well as the trivially satisfied condition $u_2 \wedge J \wedge J$, it can be shown that $u_2$ is zero, and hence we remove it from the following equations.  Using \eqref{gamma3sol} in \eqref{fullJJp} we can now consider the $(3,1)$ and $(1,3)$ components. They are
\begin{eqnarray}
\gamma_{2+}^{(2,0)} \wedge J &=& \frac{1}{2} e^{\Delta}(\overline{W}_{5} - 2 W_4 + \Theta) \wedge \Omega  \\
\;\; \textnormal{and} \;\; \gamma_{2+}^{(0,2)} \wedge J &=& \frac{1}{2} e^{\Delta}(W_{5} - 2 W_4  + \Theta ) \wedge \overline{\Omega} \;\; .
\end{eqnarray}
These may be solved for $\gamma_{2+}$ simply by contracting with $J$. We find the following,
\begin{eqnarray}
\gamma_{2+}^{(2,0)} = \frac{i}{10} e^{\Delta} ( \overline{W}_5 -2 W_4 + \Theta) \llcorner \Omega \;,
\end{eqnarray}
and the conjugate equation for $\gamma_{2+}^{(0,2)}$.

We now consider equation \eqref{floweq3}. Taking the wedge product of this equation with $J$ we obtain 
\begin{eqnarray}
\alpha_{2+} \wedge J \wedge J = e^{\Delta} \left(-W_4+ \Theta+A_{2-} \right) \wedge J \wedge J \; .
\end{eqnarray}
Because there are no primitive one forms, this tells us that,
\begin{eqnarray} \label{mr0}
\alpha_{2+} &=&e^{\Delta} (-W_4 + \Theta + A_{2-})  \;,\\
&=& e^{\Delta} (2 W_4 -2 W_{5+} +\Theta) \;.
\end{eqnarray}
In this last expression we have used equation \eqref{A2mineq}. We can also consider the wedge product of \eqref{floweq3} with $\Omega_{\pm}$. These give
\begin{eqnarray} \label{mr1}
\alpha_{1-} &=& \frac{3}{2} e^{\Delta} W_{1+} -3 e^{\Delta} A_{1+} \;\;\;\textnormal{and} \\ \label{mr2}
\alpha_{1+} &=& -\frac{3}{2} e^{\Delta} W_{1-} -e^{\Delta} A_{1-}-f e^{\Delta}\;, \\ \label{mr3}
&=& - 3 e^{\Delta} W_{1-} - \frac{15}{8} f e^{\Delta} \;.
\end{eqnarray}
Here we have used \eqref{thisguy2} and, using \eqref{mrgamma1}, equation \eqref{mr1} simplifies to $\alpha_{1-} = \frac{3}{2} \gamma_1$.  Given that $f$ is a constant, equation \eqref{mr3} is a non-trivial constraint on $\alpha_{1+}$ and $W_{1-}$. Finally, we can use \eqref{mr0}, \eqref{mr1} and \eqref{mr2} in \eqref{floweq3} to find the remaining condition
\begin{eqnarray}
\alpha_3 = e^{\Delta}( W_3 - *A_{3+} ) \;.
\end{eqnarray}
This can be solved trivially for $A_{3+}$ which is heretofore still undetermined.

Collecting everything together, and including the Bianchi identities \eqref{floweq4}, \eqref{floweq6} and form field equation of motion \eqref{floweq5}, we have the following conditions following from the flow equations.
\begin{itemize}
\item Equations from the supersymmetry variations:
\begin{eqnarray} \label{A1pluseq}
A_{1+} &=& - \frac{1}{2} e^{-\Delta} \hat{\phi}' \\
 \label{alpha3eq}
A_{3+}&=&  -*(W_3 -  e^{-\Delta} \alpha_3 ) \\
\Theta   &=& -2 W_4 +2 W_{5+} + \alpha_{2+} e^{-\Delta} \\
\gamma_1 &=& e^{\Delta} \left(W_{1+} -2 A_{1+} \right) \\
\label{gamma3eq}
\gamma_3 &=& e^{\Delta} W_{2+}  \\
\gamma_{2+}^{(2,0)} &=& \frac{i}{10} e^{\Delta}(3 W_{5+} -i W_{5-} -4 W_4 + \alpha_{2+} e^{-\Delta}) \llcorner \Omega \label{gamma2eq}
 \\\alpha_{1-} &=&\frac{3}{2} \gamma_1 \label{alpha1eq}\\
 \alpha_{1+} &=&  -3 e^{\Delta} W_{1-} - \frac{15}{8} e^{\Delta} f \label{alpha1eq2}
\end{eqnarray}
\item In addition we must satisfy the Bianchi identities $H' =dH_y$, $f'=0$ and the form field equation of motion $(* e^{-2 \hat{\phi} -\Delta} H_y)' =- d * (e^{-2 \hat{\phi} + \Delta} H ) $.
\end{itemize}

\subsection{An example of how to use this analysis} \label{case}

Let us consider the case where we are given some $SU(3)$ structure, $\Omega^{(0)}$, $J^{(0)}$,  on a six-dimensional manifold, but know nothing whatsoever about its possible deformations. We can nevertheless induce a $y$ dependent $SU(3)$ structure simply by changing an overall factor in $\Omega$ and $J$ in a compatible fashion
\begin{eqnarray}\label{parameg}
\Omega &=& a(y)^{\frac{3}{2}} \Omega^{(0)} \\
J &=& a(y) J^{(0)} \; .
\end{eqnarray}
Here $a$ is a real parameter and the power of $\frac{3}{2}$ ensures that the relationship between $J \wedge J \wedge J $ and $\overline{\Omega} \wedge \Omega$ holds for all $a$.\footnote{Note in general one could allow $a$ to depend upon the six-dimensional space too. Here we consider the simplest possible case to illustrate how to utilize our results.} Such a $y$ dependence of the $SU(3)$ structure leads to a $y$ dependence of the associated torsion classes
\begin{eqnarray}
W_1 =\frac{1}{\sqrt{a}} W_1^{(0)} \;\;,\;\; W_2 = \sqrt{a} W_2^{(0)} \;\;,\;\; W_3 = a W_3^{(0)} \;\;,\;\; W_4 = W_4^{(0)} \;\;,\;\; W_5 = W_5^{(0)}  \; .
\end{eqnarray}
Given such an explicit $y$ dependent $SU(3)$ structure we can now apply the general formalism we have just developed. Taking the $y$ derivative of $\Omega$ and $J$, we find, in terms of our general expansions in equations \eqref{jprime}-\eqref{finalomegaprime}, 
\begin{eqnarray}
\gamma_1 = \frac{2}{3} \alpha_{1-} = \frac{2}{3} \beta_{1+} &=& \frac{a'(y)}{a(y)} \\
\alpha_{1+} =\alpha_{2+} = \alpha_3 &=& 0 \\
\beta_{1-} = \beta_{2+} = \beta_3 &=& 0 \\
\gamma_{2+} = \gamma_3 &=&0  \; .
\end{eqnarray}

Given this input, we can simply use the conditions \eqref{chargesol}--\eqref{A2mineq} and \eqref{A1pluseq}--\eqref{alpha1eq2} to write down requirements on the torsion classes if a solution to the Killing spinor equations is to exist, together with the requisite fluxes in terms of the torsion classes of the $SU(3)$ structure. We find that the following constraints on the torsion classes follow immediately,
\begin{eqnarray} \label{Wegconstrs1}
W_{2+} &=& 0 \;\;, \;\; W_4 =\textnormal{exact} \;\;,\;\;  W_{5} = 2 W_4^{(0,1)} \;\;,\;\; W_{1+}^{(0)} = c_1 e^{-\Delta} \;,\label{Wegconstrs2}
\end{eqnarray}
where $c_1$ is a constant. One of the constraints on the supergravity fields is $f = - \frac{8}{5} W_{1-}$. Given that $f$ is a constant and the scaling of $W_{1-}$ given above we see that, in any non-trivial case, $f=W_{1-}=0$. Using this we then find the following solutions for the remaining supergravity fields
\begin{eqnarray}
d \hat{\phi} &=& W_4 \\
H &=& - \frac{1}{2} e^{-\Delta} \hat{\phi}' \Omega_+  + (J \llcorner W_4) \wedge J - *W_3 \\
H_y &=&   - e^{\Delta} W_{2-} \\
\Theta &=& 0 \; .
\end{eqnarray}
In this situation the domain wall dependence of the $SU(3)$ structure is simply given by
\begin{eqnarray}
a' = \sqrt{a} c_1 - a \hat{\phi}' \;.
\end{eqnarray}
Note that, due the fact that $W_4$ is independent of $y$,  $\hat{\phi}'$ is indeed only a function of $y$ and so this equation can be consistently solved. The metric \eqref{metric} can be derived explicitly, in the case of any given $SU(3)$ structure, from the solution for $\Delta$, the fact that $A=0$, and the usual formalism for deriving the metric associated to such a six manifold from the forms $J$ and $\Omega$ \cite{hitc}.To ensure that one has a good domain wall solution, it is necessary to check the pole and zero structure of the solution for $a(y)$. In addition, of course, the Bianchi identities and equations of motion for the form fields, along with any flux quantization conditions must also be satisfied.

\section{Calabi-Yau vacua with flux}

In this section we will discuss a particularly simple example which is useful to illustrate the importance of considering the form-field equations of motion and Bianchi identities in identifying solutions. We will take the six-dimensional manifold to be a Calabi-Yau threefold. That is, we shall set all of the torsion classes to zero. In doing this the equations following from preserving ${\cal N}=1/2$ supersymmetry simplify as follows 
\begin{eqnarray} \label{weasel1}
d \hat{\phi} &=&0 \;\;,\;\;
\alpha_{1+} = - \frac{15}{8}  e^{\Delta} f \;\;,\;\; A_{1-}= -\frac{7}{15} \alpha_{1+} e^{-\Delta}\;\;, \\ \nonumber
B_2 &=& B_{3+} \;=\; A_{2-} \;=\; 0  \\ \nonumber
A_{1+} &=& - \frac{1}{2} e^{-\Delta} \hat{\phi}' \;\;,\;\; B_1 = \frac{8}{15} \alpha_{1+} \;\;,\;\; A_{3+} = * (e^{-\Delta} \alpha_3) \;\;,\;\; \Theta = \alpha_{2+} e^{-\Delta} \\ \nonumber
\gamma_1 &=& -2 e^{\Delta} A_{1+} \;\;,\;\; \gamma_3 = 0 \;\;,\;\; \gamma_{2+}^{(2,0)} = \frac{i}{10} \alpha_{2+} \llcorner \Omega \;\;,\;\; \alpha_{1-} = \frac{3}{2} \gamma_1 \;\;.
\end{eqnarray}

To simplify these equations further, we must compute the $\alpha$'s, $\beta$'s and $\gamma$'s in this particular case. This is a simple exercise in special geometry (see \cite{Candelas:1990pi} for details). The holomorphic three form can be expanded in the usual basis $(\alpha_Q, \beta^Q)$ of the cohomology $H^3=H^{3,0} \oplus H^{2,1}  \oplus H^{1,2}  \oplus H^{0,3}$ on the Calabi-Yau\footnote{Our basis forms are normalized such that $\int_X \alpha_Q \wedge \beta^P = \delta^P_Q$ and $\int_X \alpha_P \wedge \alpha_Q = \int_X \beta^P \wedge \beta^Q =0$.}
\begin{eqnarray}
\Omega= {\cal Z}^Q \alpha_Q - {\cal G}_Q \beta^Q \; .
\end{eqnarray}
Here ${\cal Z}^Q$ are the homogeneous coordinates on complex structure moduli space, ${\cal G}$ is the pre-potential and ${\cal G}_Q = \partial {\cal G} / \partial {\cal Z}^Q$. Promoting the complex structure moduli to be a function of the domain wall directions, ${\cal Z} = {\cal Z}(y)$, one can compute the following parametrization of the $y$ dependence of $\Omega$
\begin{eqnarray}
 \alpha_{1-} &=&\beta_{1+}  \;\;,\;\;\alpha_{1+} = -  \beta_{1-}  \;\;,\;\; \alpha_{2+} = \beta_{2+} \;=\;0 \\
\beta_{1+} &=& \frac{1}{2} \frac{{\cal K}'}{{\cal K}} \\
\beta_{1-} &=& \frac{i}{2} \left({\cal Z}^{Q \prime} \frac{\partial}{\partial {\cal Z}^Q} - \overline{{\cal Z}}^{Q \prime} \frac{\partial}{\partial \overline{{\cal Z}}^Q} \right) \log ( {\cal K} ) \\
\beta_3+ i \alpha_3 &=& \Omega' - \overline{\beta_1} \Omega \; ,
\end{eqnarray}
where $\beta_1 = \beta_{1+} + i \beta_{1-}$ as usual and we have defined ${\cal K} =i ({\cal G}_Q \overline{{\cal Z}}^Q - \overline{{\cal G}}_Q {\cal Z}^Q )$.

To compute the $\gamma$'s we expand the K\"ahler form of the threefold in terms of a basis of $H^{1,1}$, $\omega_i$\footnote{Note that one may worry that, since what a $(1,1)$ form is can change as the complex structure varies, one might have to include $y$ dependence in the basis forms $\omega_i$ in what follows. In fact, this is not the case, essentially because all harmonic two forms on a Calabi-Yau are $(1,1)$ and so there is nothing for the $\omega$'s to vary into.}
\begin{eqnarray}
J = a^i  \omega_i \; .
\end{eqnarray}
The $a^i$ here are nothing but the real parts of the K\"ahler moduli. Promoting these to be functions of the domain wall direction, $a= a(y)$, and defining $\kappa=d_{ijk}a^i a^j a^k$, where $d_{ijk}$ are the intersection numbers of the Calabi--Yau, we find 
\begin{eqnarray}
\gamma_1 &=& \frac{1}{3} \frac{\kappa'}{\kappa} \;\;,\;\; \gamma_2 = 0 \;\;\,\;\; \gamma_{3+} = (a^{i \prime}-\frac{1}{3} \frac{\kappa'}{\kappa}a^i)\omega_i \; .
\end{eqnarray}

Using this information, the supersymmetry relations \eqref{weasel1} become the following
\begin{eqnarray} \label{phibadger}
\hat{\phi}' &=& \frac{1}{3} \frac{\kappa'}{\kappa} = \frac{1}{3} \frac{ {\cal K}'}{{\cal K}}= \frac{a^{i \prime}}{a^i} \;\; \forall i 
\\  f &=& \frac{8}{15} \beta_{1-} \;\;,\;\; \Delta = \Theta \;=\;0 \\ \label{cyhy}
H_y &=& -f J \\ \label{cyh}
H &=& -\frac{1}{6} \frac{\kappa'}{\kappa} \Omega_+ + \frac{7}{8} f \Omega_- + * \textnormal{Im} (\Omega' - \overline{\beta}_1 \Omega)\\ \label{uberbadger}
\textnormal{where} \;\; \beta_{1-} &=& \frac{i}{2} \left({\cal Z}^{Q \prime} \frac{\partial}{\partial {\cal Z}^Q} - \overline{{\cal Z}}^{Q \prime} \frac{\partial}{\partial \overline{{\cal Z}}^Q} \right) \log ( {\cal K})  = \textnormal{constant} \\ \label{badger2}
\beta_{1+} &=& \frac{1}{2} (\log {\cal K})' \; .
\end{eqnarray}

At this stage the situation looks promising. It seems like we will be able to have a ${\cal N}=1/2$ solution with non-trivial flux and domain wall dependence being present. Indeed many of the equations above obviously match those given in the analysis of \cite{Lukas:2010mf}. However, we must also study the form field Bianchi identities and equations of motion.

In the case at hand, the independent Bianchi identities and form equations of motion simplify as follows
\begin{eqnarray}
dH &=& 0 \;\;,\;\; H' = 0 \;\;,\;\; (*   e^{-2 \hat{\phi}} H_y)' = -d *(e^{-2 \hat{\phi}} H )\; .
\end{eqnarray}
The first of these clearly holds, upon using equation \eqref{cyh}, and the remaining two equations reduce to simply $H' =(e^{-2 \hat{\phi}} * H_y) ' =0$. The equation $(e^{-2 \hat{\phi}} * H_y) ' =0$ is automatically solved by equations \eqref{phibadger} to \eqref{badger2}. In the case where $f \neq 0$, which is how our analysis extends the results of \cite{Lukas:2010mf} for this example, the equation $H'=0$ is a little more long winded to analyse. Expanding in our basis, $(\alpha_P, \beta^Q)$, and setting the coefficients of the independent three forms to zero, we find $2 h^{2,1}+2$ equations determining the $y$ dependence of ${\cal Z}^Q$. In combination with equations \eqref{uberbadger} and \eqref{badger2} above, this leads to a total of $2 h^{2,1}+4$ equations for $2 h^{2,1}+2$ variables and we thus have, naively, an over-constrained system. Given this, these $y$ dependent configurations will only be present in special cases where the pre-potential allows for a solution to these equations. When this is possible will depend upon the detailed structure of ${\cal G}$ in a given case, and so we shall stop our general discussion of Calabi-Yau backgrounds here.

To conclude, as was stressed in \cite{Lukas:2010mf}, it is, in fact, possible to include flux in backgrounds where the six-dimensional manifold is a Calabi-Yau threefold. The resulting ${\cal N}=1$ theory will not have an ${\cal N}=1$ vacuum, but rather an ${\cal N}=1/2$ perturbative domain wall lowest energy state. Ideally one would like to then add non-perturbative effects in the four-dimensional description to create a stable vacuum with maximal space time symmetry. However, as was discussed for example in \cite{hetCYstable}, this can be extremely difficult in heterotic theories, and so other stabilization mechanisms may well be necessary in this case.

\section{Coset examples}
In this section we apply the general analysis of ${\cal N}=1/2$ heterotic solutions to two types of coset examples. We first recall results of heterotic compactifications on half-flat cosets \cite{Lukas:2010mf,Klaput:2011mz},  and then study a novel example, which we denote ``flipped half-flat''. For related work on heterotic compactifications on cosets, see \cite{earlycoset,zoupanosetal,nolleetal}.

Before presenting examples, let us recall a few basic facts about coset spaces (the reader is referred to the reviews \cite{cosetreview} and also to \cite{Klaput:2011mz,Koerber:2008rx} for more details).  A coset space $G/H$ allows an $SU(3)$ structure if $H$ is in $SU(3)$, and all such cosets in six dimensions are listed in \cite{Koerber:2008rx}. The geometry of a coset is specified by its structure constants, which determine the exterior algebra of its Maurer-Cartan one-forms $e^i$. Using these forms, we can construct finite bases of left-invariant forms of higher degrees. A left-invariant $SU(3)$ structure is then obtained by expanding $J$ and $\Omega$ in the bases of left-invariant two- and three forms, respectively. The constant parameters of these expansions will be constrained by the orthogonality and normalization conditions of the $SU(3)$ structure, and the requirement that $\Omega$ is complex decomposable. 

Since the exterior derivatives of the Maurer--Cartan forms are known, we can readily compute the associated torsion classes. These will also be expanded in the left-invariant forms, with coefficients determined by the parameters. In particular, $W_{1}$ is given by the parameters alone, and we thus have
\be \label{eq:cosetprop}
d W_{1} = 0 \; .
\ee

The metric on the coset can be computed using \eqref{eq:hitchmetr}. This will depend on the parameters, and requiring that it is positive definite will thus lead to further constraints. It is of great practical importance that the coset metric is explicitly known, as this allows the computation of the Hodge duals and contractions which appear in the Killing spinor equations, form equations of motion and Bianchi identities. This set of equations then leads to algebraic and differential conditions on the parameters, and thus there exist heterotic ${\cal N}=1/2$ solutions on cosets if the $SU(3)$ structure has enough parametric freedom. We now investigate this in a couple of instances.

\subsection*{Half-flat cosets}
For half-flat $SU(3)$ structure manifolds, the torsion classes fulfill (with our conventions)
\be \label{eq:hftorsion}
W_{1-} \; = \; W_{2-} \; = \; W_4 \; = \; W_5 = 0 \; .
\ee
These constraints imply that all half-flat $SU(3)$ structures share the following properties
\be \label{eq:hfprop}
d \Omega_- = 0 \; , \;
d W_3 = -\frac{3}{2} d W_{1+} \wedge \Omega_- \; , \;
d *W_{2+} = d W_{1+} \wedge J \wedge J \; .
\ee
Combining this with \eqref{eq:cosetprop}, we immediately see that $W_3$ and $*W_{2+}$ are closed on half-flat cosets.

Heterotic compactifications on half-flat manifolds have previously been studied from a four-dimensional perspective in \cite{4Dhalfflat,zoupanosetal}.  Recently, restricted half-flat cosets, which also have $W_3=0$, have been used to construct heterotic ${\cal N}=1/2$ solutions, in the case when $H=H_y = f = 0$ \cite{Klaput:2011mz}. Here, we see if we can generalize this analysis by allowing non-zero fluxes. 

Half-flat cosets share the property that $\Omega_+$ and $\Omega_-$ are expanded in separate bases of left-invariant three forms. This restricts the possible $y$-dependence that we can introduce by promoting the parameters to $y$-dependent functions. In fact, it can be shown that in all cases
\be
\alpha_{1+} = \beta_{1-} = 0 \; ,
\ee
which through \eqref{alpha1eq2} and \eqref{eq:hftorsion} implies that
\be
f = 0 \; .
\ee
The remaining Killing spinor equations are then solved by
\be \label{eq:hfflux}
H = -\frac{1}{2} e^{-\Delta} \hat{\phi}' \Omega_+ + e^{-\Delta} * \alpha_3 \; \; , \; \;
H_y =0 \; .
\ee
Hence, we see that on cosets with restricted half-flat left-invariant $SU(3)$ structure, the fluxes $f, H_y$ are set to zero. This is also true on half-flat cosets with non-zero $W_3$, as that would only modify the expression for $H$. 

By restricting the $y$-dependence of this restricted half-flat $SU(3)$ structure, also this remaining flux can be put to zero. We then reproduce the Hitchin flow solution, when the six-dimensional coset and the domain wall direction combine to form a $G_2$ holonomy manifold. 

If we insist on a non-zero $H$, we must go on to study its equation of motion and Bianchi identity. The latter reads
\be
0 = d H = -\frac{1}{2} \hat{\phi}' (e^{-\Delta} \alpha_{2+} \wedge \Omega_+ + d \Omega_+) + e^{-\Delta} \alpha_{2+}\wedge * \alpha_3 + d * \alpha_3  \; ,
\ee
where we have used that the supersymmetry equations set
\be
d\hat{\phi}=0  \; \; , \; \; \Theta=e^{-\Delta} \alpha_{2+} \; .
\ee
Moreover, the form equations of motion reduce to 
\be
H'=0  \; \; , \; \; 0 = d*H = d \left( \frac{1}{2} \hat{\phi}' e^{-2\hat{\phi}} \Omega_-  -  e^{-2\hat{\phi}}  \alpha_3 \right) = 0\; ,
\ee
where, in the second equation, we have used that $\Omega_-$ and $\alpha_3$ are expanded in the same set of closed, left-invariant three forms, in addition to $\hat{\phi}$ being closed. 

We thus find two non-trivial conditions: $H$ must be closed and $y$-independent. Solving these conditions requires information about the $y$-dependence of the coset parameters, and can only be done on a case-by-case basis. For example, in the cosets studied in \cite{Klaput:2011mz}, $\Omega_{\pm}$ has a very simple parameter scaling, and introducing $y$-dependence through the parameters only allows $\alpha_{1-}=\beta_{1+}$ non-zero. Using $\alpha_{2+} = \alpha_3=0$ in conjunction with \eqref{eq:hfflux}, it is straightforward to show that the Bianchi identity for $H$ can only be solved if $H$ is zero. Thus, a non-trivial flux is not allowed. Since $W_3$ is closed on half-flat cosets, this conclusion holds also for examples with non-zero $W_3$ if $\Omega_{\pm}$ maintains the simple parameter scaling behaviour.

\subsection*{Flipped half-flat cosets}
We now turn to a novel ${\cal N}=1/2$ heterotic domain wall coset example, based upon what we shall denote as  a ``flipped half-flat'' manifold. These cosets have
\be \label{eq:fhftorsion}
W_{1+} \; = \; W_{2+} \; = \; W_4 \; = \; W_5 = 0 \; ,
\ee
which in turn implies, using \eqref{eq:cosetprop},
\be \label{eq:fhfprop}
d \Omega_+ = 0 \; , \;
d W_3 =  0 \; , \;
d *W_{2-} =  0\; .
\ee
Thus, compared to the restricted half-flat examples discussed in the previous section, we add a non-zero $W_3$, and thus an extra term to $H$, and turn on imaginary, and not real, $W_{1-}$ and $W_{2-}$. We will now investigate if we can find non-trivial fluxes that solve all the heterotic ${\cal N}=1/2$ conditions.

To be concrete we focus on the coset $\frac{SU(2)^2}{U(1)} \times U(1)$ which allows the following flipped half-flat left-invariant $SU(3)$ structure\footnote{This is not the most general left-invariant expansion of $J$ and $\Omega$, but it will suffice to allow a non-zero $W_{1-}, W_{2-}$ and $W_3$. This coset, but with a half-flat $SU(3)$ structure, has been studied in \cite{Caviezel:2008tf} in the context of type II compactifications.} 
\begin{align}
J &=  b_1 ( e^{15} + e^{24} ) + b_2 e^{36} \ , \\ \nn
\Omega &= 
d_1 e^{126} + d_2 (-e^{134} + e^{235}) + d_3 e^{456}
+ \frac{i d_2}{\sqrt{d_1 d_2^2 d_3}} \left[
-d_1d_2 e^{123} - d_1 d_3 (-e^{146}+e^{256}) - d_2 d_3 e^{345}
\right]\ .
\end{align}
Here $e^{ij} = e^i \wedge e^j$, $b_i$ and $d_i$ are real parameters, and the coefficients of $\Omega_-$ have been chosen to ensure complex decomposability of $\Omega$.  Orthogonality \eqref{eq:orth} is automatic for this example, whereas the normalization condition \eqref{eq:norm} imposes
\be
b_2 =  \frac{\sqrt{d_1 d_2^2 d_3}}{b_1^2}  =  \frac{C}{b_1^2}\; ,
\ee
where we introduce the shorthand $C=\sqrt{d_1 d_2^2 d_3}$. Metric positivity further requires that the following parameter combinations are all strictly positive:
\be
b_1 d_1 d_2 > 0 \; , \;
b_1 d_2 d_3 > 0 \; , \;
d_2^2  > 0 \; , \; 
d_1 d_3  > 0 \; ,
\ee
The above computations, and the subsequent analysis of the torsion classes, is simplified by using the symbolic computer program \cite{Bonanos}. Using this, it is easily shown that the non-zero torsion classes are
\begin{eqnarray} \label{cosettorsion}
W_{1-}&= & \frac{ (d_1 + d_3 )d_2^3 + 2 b_1^3 C}{6 C b_1^2} \nn \\
W_{2-}&= & \frac{b_1-2W_{1-}d_2}{b_1^2 d_1}  \left( b_1^3 (e^{15}+ e^{24}) -2 C e^{36}\right)
\nn \\
W_3 &=& -\frac{3}{2} W_{1-} \Omega_+ -
\frac{1}{b_1^2} \Big[
C(e^{126}+e^{456}) + b_1^3 (e^{134}-e^{235})
\Big]
\; .
\end{eqnarray}

Promoting the parameters to $y$-dependent functions, and comparing with the general $y$-dependence  \eqref{omegaprime} for $\Omega'$ and \eqref{jprime} for $J'$, then results in
\begin{align}
\label{cosetydep1}
\beta_{1+} &= \alpha_{1-} = \frac{1}{4} \left(
\frac{d_1'}{d_1} + 2 \frac{d_2'}{d_2} + \frac{d_3'}{d_3} 
\right)  \nn \\
\alpha_3 &= \Omega_-' - \alpha_{1-} \Omega_- \nn \\
\beta_3 &=  \Omega_+' - \beta_{1+} \Omega_+
\nn \\ 
\gamma_1 &= \frac{1}{6} \left(
\frac{d_1'}{d_1} + 2 \frac{d_2'}{d_2} + \frac{d_3'}{d_3} 
\right) \nn \\
\gamma_{3} &= 
\frac{1}{b_1^3} \left(b_1'  - b_1 \gamma_1 \right)
\left[b_1^3 (e^{15}+e^{24}) - 2 C e^{36} \right]\; , \;
\end{align}
with all other $\alpha_i, \beta_i, \gamma_i$ being zero. It is straightforward to check that $\alpha_3, \beta_3$ and $\gamma_3$ fulfill their respective primitivity conditions. 

With this information at hand we turn to the supersymmetry variations \eqref{chargesol}--\eqref{A2mineq} and \eqref{A1pluseq}--\eqref{alpha1eq2}. The absence of $W_{2+}$ sets $\gamma_3$ to zero:
\be
W_{2+}=0 \implies \gamma_3 = 0 \Longleftrightarrow \left(\ln b_1 \right)' =\gamma_1 \; .
\ee
Moreover,  \eqref{alpha1eq}  is automatically satisfied. The remaining equations are then solved by
\begin{align}
\hat{\phi}' &= \gamma_1   \; \; , \; \;
\Theta= 0\; \; , \; \;
d \hat{\phi}  =  0 
\; ,
\end{align}
and
\begin{align}
f &=- \frac{8}{5} W_{1-}
\nn \\
H &= -\frac{1}{2}e^{-\Delta} \hat{\phi}' \Omega_+
-\frac{1}{10} W_{1-} \Omega_-
-* (W_3 - e^{-\Delta} \alpha_3) \nn \\
H_y &= -\frac{2}{5} e^{\Delta}W_{1-} J -  e^{\Delta} W_{2-} \; .
\end{align}

In addition to the Killing spinor equations, we must show that the form equations of motion \eqref{formeom}  and Bianchi identities \eqref{eq:bi} are satisfied. As we have seen before, with $\alpha_{1+}=0$ the Bianchi identity for $f$ requires that $W_{1-}' = 0$ is a constant, so\footnote{The Bianchi identity for $H$ can be shown to be incompatible $W_{1-}$ being zero.}
\be
\left(\frac{ (d_1 + d_3 )d_2^3 + 2 b_1^3 C}{6 C b_1^2}  \right)' = 0  \; .
\ee
We can solve this for, say $d_3'$. Note that there is still some $y$-dependence left after this choice has been made, in particular $W_{2-}'$ and $W_3'$ remain non-zero.

We now turn to the form equations of motion. The first of these reduces to 
\be
0 = d(* e^{-2 \hat{\phi}- \Delta} H_y)  
\; \Longleftrightarrow \; 
0 = d * W_{2-}
\ee
which is automatically satisfied  on flipped half-flat cosets, see \eqref{eq:fhfprop}. The second equation of motion is 
\begin{align} \label{eq:coseteom2}
(* e^{-2 \hat{\phi} -\Delta} H_y)' &= - d * (e^{-2 \hat{\phi} + \Delta} H ) \nn \\
 d \alpha_3 
&=
\left(W_{2-}' - \frac{1}{2} \gamma_1 W_{2-} \right) \wedge J 
+\frac{1}{2} \gamma_1 W_{1-} J \wedge J \; ,
\end{align}
where we have used \eqref{eq:hfprop}. This condition is not automatic. A simple way of solving it is to put all $y$-dependence to zero.

The $H$-flux Bianchi identity is independent of the $y$-dependence we choose, and reads
\be \label{eq:cosetbi1}
0 = d H = -\frac{1}{10} W_{1-} (W_{1-} J \wedge J + W_{2-} \wedge J) - d * W_3 \; ,
\ee
where we have used that $* \alpha_3$ is expanded in the same closed left-invariant forms as $\Omega_+$. This constraint can be expanded in the basis of left-invariant forms, and we get one polynomial condition for each non-zero expansion coefficients. These conditions thus define an algebraic ideal, and algebraic geometry methods can be used to analyse their solution space. There are solutions to this equation that are compatible with the Bianchi identity for $f$. 

The second Bianchi identity reads
\begin{align}
\label{eq:cosetbi2}
d H_y &= H' \Longleftrightarrow \nn \\
e^{\Delta} \left(\frac{3}{5} W_{1-}^2 \Omega_+ - \frac{2}{5} W_{1-} W_3- dW_{2-} \right) &= 
-\frac{1}{2} e^{-\Delta}  \left(\gamma_1' + \frac{3}{2} \gamma_1^2 \right) \Omega_+ 
-\frac{3}{20} W_{1-} \gamma_1 \Omega_{1-}
\nn \\
&- \frac{1}{2} e^{-\Delta} \gamma_1 \beta_3
- \frac{1}{10} W_{1-} \alpha_3
-\left[ *(W_3 - e^{-\Delta} \alpha_3) \right]'   \; .
\end{align}
If we set the $y$-dependence to zero, this equation requires that $d W_{2-}$ is a linear combination of $\Omega_+$ and $W_3$.

The differential constraints \eqref{eq:coseteom2} and \eqref{eq:cosetbi2} are the most difficult to solve. Expanding them in the left-invariant form basis results in an over constrained system, that naively lacks solutions. This naive observation turns out to be true. For trivial $y$-dependence, the conditions are just polynomial equations in the parameters. These can be analysed, using, for example, the Mathematica package {\it Stringvacua} \cite{Gray:2008zs}, with the result that there are no $y$-independent solutions. In the general case, with non-trivial $y$-dependence, we can also show that there are no solutions by eliminating the $y$-dependent parameters, until the remaining constraints are incompatible with the Bianchi identity for $H$. 

In conclusion,  we can find fluxes that solve the Killing spinor equations and one of the form equations of motion on flipped half-flat cosets. However, in the specific example we have studied, the parametric freedom is not enough to satisfy all Bianchi identities, including the one for $H_y$. Whether there are other flipped half-flat coset examples satisfying all conditions remains an open question.

\section{A toric example}

We now turn to our final example geometry, which is an $SU(3)$ structure on a smooth, compact, toric variety (SCTV) with all torsion classes non-zero. SCTVs can be described as symplectic quotients of $\mathbb{C}^n$, and inherit a complex structure, a metric and a real, closed two form $J_{FS}$ from this covering space. In addition, it was shown in \cite{Larfors:2010wb} that toric spaces can admit an $SU(3)$ structure defined by a real two form $J$ and a complex decomposable three form $\Omega$:
\begin{eqnarray}
\label{sctvsu3}
J&=&\a j - \frac{i\b^2}{2}K\wedge \bar{K}\\
\Omega &=& e^{i \g} \a\b  \bar{K}\wedge\omega
~.
\end{eqnarray}
Here $\a$, $\b$, $\g$ are nowhere-vanishing, real functions on the toric variety, and will be allowed to depend also on $y$ in what follows. The two form $j$ is related to the inherited two form $J_{FS}$ by 
\be
j=J_{FS}-\frac{i}{2}K\wedge \bar{K}\;.
\ee 
The one-form $K$ is defined on $\mathbb{C}^n$ and obeys certain properties which are described in detail in \cite{Larfors:2010wb}. This one-form and the two form $\omega$, which is built out of $K$ and ambient space quantities, are not independently well defined on the quotient manifold. However, $K \wedge \overline{K}$ and $\overline{K} \wedge \omega $ are and hence the above expressions make sense. More details on the construction of this type of $SU(3)$ structures can be found in \cite{Larfors:2010wb,Larfors:2011zz} and in Appendix \ref{ap:sctv}.

Generally, the torsion classes of these SCTV $SU(3)$ structures are all non-zero, and have to be computed on a case by case basis. For concreteness, we will therefore focus on one particular example below. However, there is one common feature, namely  a limit in the parameter space spanned by $\a, \b$ and $\g$ where the torsion classes simplify. When $\a=-\b^2$ it is readily shown that 
\be
J \propto J_{FS}\ , \mbox{ if } \a=-\b^2
\ee
and hence the torsion classes simplify to 
\be
W_1 = W_3 = 0, \; \; \; \; \;  W_2,\; W_4 = d \a, \; W_5 \mbox{ non-zero} , \mbox{ if } \ \a=-\b^2 . 
\ee
Any SCTV $SU(3)$ structure can be made $y$-dependent by allowing the parameters $\a, \b$ and $\g$ to depend on $y$.\footnote{One could also introduce another simple form of $y$ dependence by allowing the parameters which appear in the moment map conditions of the symplectic quotient to vary in the domain wall direction. The authors will pursue a more complete analysis of such vacua in a future publication.} This leads to $y$-derivatives
\begin{align}
J' &=  \a' j - i \b \b' K \wedge \bar{K} 
 \nonumber \\
\Omega' &= \left((i \g) + \ln [\a \b] \right)' \Omega
\ ,
\end{align}
which can be compared with the general expansion \eqref{omegaprime} for $\Omega'$ and with \eqref{jprime} for $J'$. The result is
\begin{eqnarray}
\label{sctvydep1}
\alpha_{1-} &=& \beta_{1+} = \left(\ln \a\b \right)'  \nn \\
\alpha_{1+} &=& -\beta_{1-} = \g'  \nn \\
\gamma_1 &=& \frac{2}{3} \left(\ln \a \b \right)' \nn \\ \label{mrgamma2}
\gamma_{3} &=& 
a \left((\ln \a)' - \frac{2}{3}(\ln \a\b)' \right) j 
- \frac{i \b^2}{2}\left(2(\ln \b)' - \frac{2}{3}(\ln \a\b)'\right) K \wedge \bar{K}
\; , 
\end{eqnarray}
with all other $\alpha_i, \beta_i, \gamma_i$ zero.\footnote{That $\gamma_{2+}=0$ is a consequence of the fact that $j \wedge \omega = 0$ \cite{Larfors:2010wb}.} As a check of this result one can verify that $\gamma_{3}$ is a primitive form.  

\medskip

Let us now investigate whether the supersymmetry variations \eqref{chargesol}--\eqref{A2mineq} and \eqref{A1pluseq}--\eqref{alpha1eq2} are fulfilled by the SCTV example of Appendix \ref{ap:sctv}, which is a a class of toric $\mathbb{CP}^1$ fibrations whose $SU(3)$ structure was first constructed in \cite{Larfors:2010wb}. The associated torsion classes are given in full generality in \eqref{sctvtorsions}. They are all non-zero and $W_1, W_2, W_5$ are complex. Moreover, $W_5$ is $\partial$-exact and $W_4$ can be made exact for certain parameter choices:
\be
 \a = \hat{C} \b^2 \; \; \Longrightarrow \; \; \; W_4 \; \mbox{ exact } ,
\ee
where $\hat{C}$ is a constant. We note that this relationship, used in conjunction with equation \eqref{mrgamma2}, implies that $\gamma_3 =0$.

Combining this result with \eqref{sctvydep1} and \eqref{gamma2eq} implies that
\begin{equation}
\overline{W}_{5} - 2 W_4^{(0,1)} = 0 \ .
\end{equation}
This condition is solved, using $W_4$ and $W_5$ in \eqref{sctvtorsions}, by
\begin{equation}
\g = \g(y)  \; \; \mbox{and} \; \b = \frac{C(y)}{p}\ e^{- 2\hat{C} F} \ .
\end{equation}
Here, $C(y)$ is an undetermined real function of $y$, whereas $p$ and $F$ are real functions of the SCTV coordinates that are defined in Appendix \ref{ap:sctv}.

Since $\gamma_{3} = 0$, which through \eqref{gamma3eq} implies that $W_{2}$ must be imaginary, we have to choose
\be
\g (y) = n \pi \; ,\; n \in \mathbb{Z} \ .
\ee
This choice also puts $W_{1+}=0$, as well as $\alpha_{1+} = \beta_{1-} = 0$. Thus, the $y$-dependence of the $SU(3)$ structure is restricted to the simple scaling of $J$ and $\Omega$ discussed in section \ref{case}. In particular, the $y$-dependence of the torsion classes is completely determined:
\begin{eqnarray} \label{toothischaptoo}
W_1 =\frac{1}{C(y)} W_1^{(0)} \;\;,\;\; W_2 = C(y) W_2^{(0)} \;\;,\;\; W_3 = C^2(y) W_3^{(0)} \;\;,\;\; W_4 = W_4^{(0)} \;\;,\;\; W_5 = W_5^{(0)}  \; .
\end{eqnarray}
The rest of the equations \eqref{chargesol}--\eqref{A2mineq} and \eqref{A1pluseq}--\eqref{alpha1eq2} are then solved by
\begin{align} \label{sctvsol}
d \hat{\phi} &= W_4\; \; , \; \; \hat{\phi}' = 2 (\ln C(y))' \\
\Theta &= -2 W_4 + 2 W_{5+} = 0 \implies \Delta \mbox{ is constant} \nn \\ \nn 
f &= -\frac{8}{5} W_{1-} \\
H &= -e^{-\Delta} (\ln C(y))' \Omega_+ + \frac{1}{10} W_{1-} \Omega_- + W_{5-} \wedge J -*W_3 \nn \\ \nn 
H_y &= e^{\Delta} \left( -\frac{2}{5} W_{1-} J -  W_{2-} \right) \; .
\end{align}
Interestingly, we notice that it is possible to put $\a=-\b^2$ and still maintain this solution. This is the simplifying limit mentioned above where only $W_2$, $W_4$ and $W_5$ remain non-zero, and hence the fluxes are simplified too. Since all SCTV $SU(3)$ structures have this limit, we see that any such variety  allows a heterotic ${\cal N}=1/2$ solution, given that $W_5$ is $\partial$-exact (note that $W_4$ is automatically exact in this limit). Since the class of six-dimensional smooth compact toric varieties is large, we can thus expect to find more examples.

Let us now turn to the form field equations of motion, given in \eqref{formeom}. After inserting the fluxes in \eqref{sctvsol}, we find that the two equations become
\begin{align} \label{sctveom1} 
d(W_{2-}\wedge J) &= 2 W_4 \wedge W_{2-} \wedge J + \frac{1}{5} dW_{1-} \wedge J \wedge J\\ \nn
-\frac{4}{5} (\ln C)' W_{1-} J \wedge J &= e^{\Delta}( d W_3 - 3 W_4 \wedge W_3) \\ \nn
&+ \left[-(\ln C)' W_{4} \; \;  + \frac{e^{\Delta}}{10} W_{1-} W_{5-}\right] \wedge \Omega_- \\ \nn
&+\left[  - (\ln C)' W_{5-}+ \frac{e^{\Delta}}{10} W_{1-}(14  W_{4}+ d\ln W_{1-} )\right] \wedge \Omega_+
 \; .
\end{align}

These expressions simplify considerably if we take $\a=-\b^2$, reducing to 
\begin{align} \label{sctveom1b} 
d(W_{2-}\wedge J) &= 2 W_4 \wedge W_{2-} \wedge J \\ \nn
0 &= 
(\ln C)'  \left(W_{5+} \wedge \Omega_- + W_{5-} \wedge \Omega_+\right)
 \; ,
\end{align}
where we have used that $W_{5+}=W_4$. Since $W_5$ is the holomorphic $(1,0)$ form, it is straightforward to show that the last of these two equations is satisfied. Moreover, it can be shown that the first equality follows from $d^2 \Omega = 0$. Thus, the form equations of motion can be satisfied by setting $\a=-\b^2$ in this example. This is promising for a generalization of this solution to other SCTVs, for the reasons discussed underneath equation \eqref{sctvsol}.

Finally, we turn to the Bianchi identities of the system, and focus on the limit  $a=- b^2$. In this limit, the condition $f' =0$ is automatic, given that $W_1$ is set to zero. The remaining Bianchi identities then imply the conditions
\begin{align}
0 &=  e^{-\Delta} (\ln C)' d \Omega_+ + \left(d W_{5-}+ W_{5-} \wedge W_{5+} \right) \wedge J \\ \nn
\ -e^{\Delta} d W_{2-}  &= e^{-\Delta} \left((\ln C)'' + 3 (\ln C)' \right) \Omega_+ + 2 (\ln C)' W_{5-} \wedge J
\ .
\end{align}
Even in this simplified case, these conditions are prohibitively difficult to verify. The problem lies in computing various contractions when calculating the torsion classes, which requires the use of the metric on the SCTV. While the relevant metric is know, the expression is too complicated to be of practical use. Consequently, our analysis remains inconclusive regarding the Bianchi identities.

\section{Conclusions}

In this paper we have analysed four-dimensional domain wall solutions of the Neveu--Schwarz sector of heterotic string theory compactified on a manifold admitting an $SU(3)$ structure. These ${\cal N}=1/2$ supersymmetric solutions are to be interpreted as the perturbative ground states of the four-dimensional ${\cal N}=1$ theories obtained by compactifying heterotic theories on the $SU(3)$ structure space in question.

In pursuing this work we are extending previous results of Lukas and Matti in \cite{Lukas:2010mf} in several regards. Firstly, we include two new fluxes that were omitted in this earlier work. This allows more general torsion classes, and thus significantly extends the set of compact geometries relevant for heterotic compactifications. Secondly, we provide explicit expressions for the supergravity fluxes in terms of the torsions classes and other data specifying the $SU(3)$ structure. This analysis is carried out in complete generality in section 4, with the main results being summarized in equations \eqref{chargesol}-\eqref{A2mineq} and \eqref{A1pluseq}-\eqref{alpha1eq2}. These results can be used for any construction of an $SU(3)$ structure on a six-dimensional manifold to find the constraints for a solution to the Killing spinor equations, and the values that the supergravity fields will take in vacuum. There is one key aspect of obtaining a solution to these systems where having an explicit expression for the fluxes is of paramount importance. In addition to the Killing spinor equations, one must check that the form-field equations of motion and Bianchi identities are satisfied. An explicit expression for the fluxes makes such an analysis possible. Indeed, we find in many cases that the constraints imposed by the Bianchi identities on the fluxes are rather restrictive and often obstruct the existence of a solution.

To both illustrate the power of the analysis in section 4, and the importance of solving the Bianchi identities and equations of motion for the form fields (rather than just the Killing spinor equations alone) we then provide a series of examples of the application of our analysis.

We began by considering the case of a completely general $SU(3)$ structure where we allowed only an overall scale to vary in the domain wall direction. We then moved on to the more complicated example of Calabi-Yau compactifications which are not deformed to torsion-full $SU(3)$ structures in the presence of flux \cite{Lukas:2010mf}. We have also discussed compactifications based upon coset manifolds \cite{Klaput:2011mz,zoupanosetal} and smooth compact toric varieties \cite{Larfors:2010wb} which admit $SU(3)$ structures. In both cases, we have provided both some general analysis, in as far as this is possible, as well as a detailed examination of a particular example.

There are many future directions which can be pursued using the results in this paper as a starting point. These include more detailed studies of the smooth compact toric varieties in a heterotic context, some aspects of which the authors plan to study in upcoming publications. One could also think of performing a similar analysis to the one presented here in type II string theory, and in particular, including the complications of the Ramond-Ramond sector. Finally, we should point out that, in this paper, we have worked to lowest order in $\alpha'$. It would be of great interest to include first order effects in this expansion for two reasons. Firstly, this is the order at which gauge fields  and five branes enter the heterotic action.  Secondly, the corrections to the Bianchi identities which occur at first order in $\alpha'$ might lead to a softening of the strong constraints on the existence of solutions which we see coming from these equations. Of course one must be extremely cautious in analysing situations where terms at one order in an expansion are balancing those at another - one must ensure that the remaining terms of higher order can still be neglected. However, such a generalization of the analysis presented here remains an intriguing possibility.

\vskip 0.5cm
\noindent
{\bf Acknowledgments\\}
The authors would like to thank Lara Anderson, Johannes Held, Michael Klaput, Andre Lukas, Cyril Matti, Xenia de la Ossa and Dimitrios Tsimpis for useful discussions. The research of M.~L. and D.~L.  was supported by the Munich Excellence Cluster for Fundamental Physics ``Origin and the Structure of the Universe''.  J.G. would like to acknowledge support by the NSF-Microsoft grant NSF/CCF- 1048082. J.~G. and D.~L. thank the Simons Center for Geometry and Physics, and J.~G. and M.~L. thank the Isaac Newton Institute for Mathematical Sciences, as part of the programme on the Mathematics and Applications of Branes in String and M-theory, for hospitality while part of this work was being completed.

\newpage
\begin{appendix}

\section{Conventions and useful relations}\label{app:conv}

In this appendix we lay out the conventions we have used in dealing with forms throughout the rest of the paper.\footnote{References with useful formulas include \cite{Lust:2004ig}; however, our notation differs from theirs in some respects.}  We define the components of a $p$-form as follows
\be
A = \frac{1}{p!} A_{m_1... m_p} d z^{m_1} \wedge ... \wedge d z^{m_p} \ .
\ee
The components of a wedge product are then given by,
\be
(A \wedge B)_{m_1...m_{p+q}} = \frac{(p+q)!}{p!q!} A_{[m_1... m_p} B_{m_{p+1}... m_{p+q}]}  \; ,
\ee
and the Hodge star operation takes the form,
\be
* A = \frac{1}{p! (d-p)!} \epsilon_{n_1 ... n_{d-p} m_1 \ldots m_p} A^{m_1... m_p} d z^{n_{1}} \wedge ... \wedge d z^{n_{d-p}} \;.
\ee
We define the contraction of a $q$-form with a $p$-form ($p > q$) by
\be
B \llcorner A = \frac{1}{(p-q)!} B^{m_1 ... m_q} A_{m_1 ... m_p} d z^{m_{q+1}} \wedge ... \wedge d z^{m_p} \ .
\ee
It is also useful to have in hand one ``higher order" operation, obtained by combining the above definitions. We have, for the Hodge star of a wedge product:
\be \label{eq:hodgewedge}
* (A \wedge B) = \frac{1}{p!} A \llcorner (* B) \ ,
\ee

For $SU(3)$ structure manifolds our conventions are such that the associated two and three forms obey
\be \label{eq:norm}
J \wedge J \wedge J = \frac{3}{4} i \Omega \wedge
  \bar{\Omega} \; , \; \mbox{and} \; *1 = \frac{1}{3!} J \wedge J \wedge J \; .
\ee

The Hodge duals of $J$ and $\Omega$ are 
\begin{eqnarray}
* \Omega_+ &=& - \Omega_- \\
* \Omega_- &=& \Omega_+ \\
* J &=& \frac{1}{2} J \wedge J \\
* (J \wedge J) &=& 2 J \; ,
\end{eqnarray}
where the $\pm$ subscripts denote real and imaginary parts of the associated form.

\vspace{0.2cm}

An Hermitian metric can be constructed from $J$ and $\Omega$ \cite{hitc} by 
\be \label{eq:hitchmetr}
g_{mn} = \mathcal{I}_m{}^l J_{ln} \ ,
\ee
where the complex structure $\mathcal{I}_m{}^l$ is defined by $\Omega_+$ \cite{hitc} and acts on a holomorphic one-form $\rho$ as 
\be
\mathcal{I}_n{}^m \rho_m = i \rho_n \ .
\ee
This implies that
\be
J^m{}_n = g^{mk} J_{kn} = - \mathcal{I}^m{}_n = \mathcal{I}_n{}^m \ ,
\ee 
which we will use in computing contractions with $J$. For example, if $\rho$ is holomorphic 1-form, then
\be \label{eq:holcontr}
\rho \llcorner J = \rho^m J_{mn} dz^n =   i \rho \ .
\ee
Similarly $\rho^* \llcorner J = - i \rho^*$.

\subsection{Useful identities for $(p,q)$-forms}\label{mrappy}

\subsubsection*{(1,0)-forms}
The formulae and definitions of the previous subsection can be used to show that any holomorphic 1-form $\rho$, with real and imaginary part $\rho_{\pm}$, fulfils
\begin{eqnarray} 
* \rho &=& \frac{i}{2} \rho \wedge J \wedge J \\
* (\rho_{\pm}) &=& \mp \frac{1}{2} \rho_{\mp} \wedge J \wedge J \\
* (\rho \wedge J) &=& i \rho \wedge J \\
* (\rho_{\pm} \wedge J) &=& \mp \rho_{\mp} \wedge J \\
* (\rho_{\pm} \wedge J \wedge J) &=& \mp 2 \rho_{\mp} \ .
\end{eqnarray} 

\subsubsection*{(1,1)-forms}
Using our relations and conventions given at the start of this section, we see that for  a primitive (1,1)-form $W_2$
\begin{eqnarray} 
W_2 \llcorner J &=& 0 \\
* W_2 &=& - W_2 \wedge J \ .
\end{eqnarray} 

\subsubsection*{(2,0)-forms} 
For holomorphic one-forms $\rho, \sigma$ 
\begin{eqnarray} \label{eq:hodge20}
* (\rho \wedge \sigma) &=& - 2 \rho \wedge \sigma \wedge J  
\ ,
\end{eqnarray} 
where the Hermiticity of the metric has been used.

A particularly interesting type of (2,0)-forms is given by contractions of the holomorphic top-form
\be \label{eq:omega}
\omega = \overline{\rho} \llcorner \Omega = \frac{1}{2} \overline{\rho}^{m} \Omega_{mnp} dz^n \wedge dz^p\ ,
\ee
where $\rho$ is a (1,0)-form. The Hodge dual of such a form is given by
\be
*\omega = *(\overline{\rho} \llcorner \Omega) = -i \overline{\rho} \wedge \Omega \ .
\ee
Similarly
\be
*\omega_+ = \frac{i}{2} ( \rho \wedge \overline{\Omega}- \overline{\rho} \wedge \Omega ) = \rho_{+} \wedge \Omega_{-}  - \rho_{-} \wedge \Omega_{+}\ .
\ee
Furthermore, one can show that 
\be
\omega \wedge \overline{\Omega} = (\overline{\rho} \llcorner \Omega) \wedge \overline{\Omega}  =  \overline{\rho} \llcorner (\Omega \wedge \overline{ \Omega}) =
-\frac{4}{3} i \overline{\rho} \llcorner (J \wedge J \wedge J) =
4 \overline{\rho} \wedge J \wedge J  \ ,
\ee
and similarly $\overline{\omega} \wedge \Omega = 4 \rho \wedge J \wedge J$.

\section{Torsion classes of a toric example}\label{ap:sctv}

Toric varieties can be described as symplectic quotients of $\mathbb{C}^n$. We denote the coordinates on $\mathbb{C}^n$ by $z^i$, where $i=1,...,n$, and consider a $U(1)^s$ action given by:
\begin{equation}
\label{sctvu1}
z^i\longrightarrow e^{i\varphi_a\tilde{Q}^a_i}z^i~,
\end{equation}
where $\tilde{Q}^a_i,~ a=1,\dots s$ are the $U(1)$ charges of $z^i$. A toric variety of real dimension $2d$  can then be  defined 
as the quotient
\begin{equation}
\label{sctvquot}
\mathcal{M}_{2d}=\{
z^i\in\mathbb{C}^n | \sum_{i=1}^n\tilde{Q}^a_i|z^i|^2=\tilde{\xi}^a
\}/U(1)^s~,
\end{equation}
where $d=n-s$ and the $U(1)^s$ act as in (\ref{sctvu1}). 
It can be shown that the Fayet-Iliopoulos parameters $\tilde{\xi}^a$  
are in fact the K\"{a}hler moduli of  $\mathcal{M}_{2d}$ (see e.g. \cite{Denef:2008wq}).

The toric $\mathbb{CP}^1$ bundle that is studied in this paper is a symplectic quotient in $\mathbb{C}^6$ specified by the moment maps
\begin{align}
\label{sctvmmaps}
|z^2|^2+|z^4|^2-2|z^5|^2&=\widetilde{\xi}^1\\ \nn
|z^1|^2+|z^3|^2-2|z^6|^2&=\widetilde{\xi}^2\\ \nn
|z^5|^2+|z^6|^2&=\widetilde{\xi}^3 \ .
\end{align}
Given this information, the forms $J$ and $\Omega$ in \eqref{sctvsu3} that specify $SU(3)$ structure can be constructed by projecting and restricting forms in $\mathbb{C}^6$ to the symplectic quotient. The procedure is described in \cite{Larfors:2010wb} (further details regarding the complex decomposability of the holomorphic three form can be found in \cite{Larfors:2011zz}). 

The torsion classes of this toric $\mathbb{CP}^1$ fibration  can be computed using the symbolic computer program \cite{Bonanos}, and were presented for constant $\a, \b$ and $\g$ in \cite {Larfors:2010wb}. In full generality, they are 
\begin{align}
\label{sctvtorsions}
&W_1= -i e^{i\g} \ \frac{2p (\a+\b^2)}{3 \a \b} \nn\\
&W_2 = i e^{i\g} \ w_2 \quad , \quad \mbox{where } w_2 \mbox{ is real and non-zero} \nn \\
&W_3 = -\frac{ (\a+\b^2)}{\b^2} d F   \wedge \left(\a j + i \frac{\b^2}{2} K \wedge \bar{K} \right) + \left(\chi - \frac{1}{4} (J \llcorner \chi )\wedge J \right)
\nn\\
&W_4 =  \frac{ (\a+\b^2)}{\b^2} d F + \frac{1}{4} J \llcorner \chi
\nn\\
&W_{5}=\partial \left( 2 F - \ln p  +  \ln (\a \b) - i  \g \right) ~,
\end{align}
where $\partial$ denotes the holomorphic (with respect to $J$) part  of the exterior derivative $d$.\footnote{The expression for $W_2$ is not written out since a simple, presentable, form of it is not known; what is computed using \cite{Bonanos} is $W_2 \wedge J$. We remind the reader that an explicit expression for $K$ can be found in \cite{Larfors:2010wb}.} Moreover, $p$ is a real, gauge-invariant, nowhere-vanishing function:
\begin{align}
&p = \frac{\sqrt{p_1}}{p_2} \quad \quad \mbox{where} \nn \\
&p_1 =  \left( |z^5|^2 + |z^6|^2\right) \left(|z^2|^2+|z^4|^2 \right)\left(|z^1|^2+|z^3|^2 \right) +
4|z^5 z^6|^2 \sum_{i=1}^{4} |z^i|^2, 
\nn\\
&p_2 = |z^5|^2 \left(|z^2|^2+|z^4|^2 \right)+
|z^6|^2 \left(|z^1|^2+|z^3|^2 \right) \; ,
\end{align}
and $F$ is a real function 
\begin{equation}
F = \ln p_2 + 2\frac{ -\tilde{\xi}^1 + \tilde{\xi}^2 }{f_1}
 \mbox{Arctan}(f_2/f_1)  ~,
\end{equation}
where
\begin{align}
f_1 &= \sqrt{-(\tilde{\xi}^1)^2-(\tilde{\xi}^2)^2+8\tilde{\xi}^2\tilde{\xi}^3+16(\tilde{\xi}^3)^2+2\tilde{\xi}^1(\tilde{\xi}^2+4\tilde{\xi}^3)},
\nonumber 
\\
f_2 &= \tilde{\xi}^1 - \tilde{\xi}^2 - 4\tilde{\xi}^3 + 8 |z^5|^2,
\end{align}
are gauge-invariant functions. Note that the $f_i$ can be zero, negative or imaginary also when we restrict the K\"ahler moduli $\tilde{\xi}^i$ to lie in the K\"ahler cone (i.e. be larger than zero). 

That $W_3$ is a primitive form follows from the fact that 
\be
d F \wedge j \wedge j = 0 \; ,
\ee
as can easily be verified. 

The three form $\chi$ that contributes to $W_3$ and $W_4$ is given by
\begin{align}
\chi &= - d \a \wedge j + i \b d \b \wedge K \wedge \bar{K} \nonumber \\
&=d \ln(\a) \wedge J + i \frac{\b^2}{2}(2 d \ln(\b) - d \ln(\a) ) \wedge K \wedge \bar{K} \ .
\end{align}
When $\a \propto \b^2$, where the proportionality coefficient is constant in the toric variety but not necessarily along the domain wall direction, we have $\chi = d \ln \a \wedge J$. This lacks a primitive piece, and so does not contribute to $W_3$. In this limit, $W_4$ is exact, and hence fulfills one of the necessary conditions for a heterotic vacuum. 

\end{appendix}

 \end{document}